\documentclass[unsortedaddress,aps,preprint]{revtex4}
\usepackage{graphicx}
\usepackage{dcolumn}
\usepackage{amsmath}
\usepackage{amssymb}
\usepackage{amsfonts}
\usepackage{bm}
\usepackage{color}
\usepackage[normalem]{ulem}
\usepackage{hyperref}
\hypersetup{colorlinks=true,linkcolor=blue,urlcolor=magenta}
\newcommand{\be}{\begin{equation}}
\newcommand{\ee}{\end{equation}}
\newcommand{\ba}{\begin{eqnarray}}
\newcommand{\ea}{\end{eqnarray}}

\begin{document}
\title{Ground state of weakly repulsive soft-core bosons on a sphere}
\author{Santi Prestipino$^1$\footnote{Corresponding author. Email: {\tt sprestipino@unime.it}} and Paolo V. Giaquinta$^1$\footnote{Email: {\tt paolo.giaquinta@unime.it}}}
\affiliation{$^1$Universit\`a degli Studi di Messina,\\Dipartimento di Scienze Matematiche e Informatiche, Scienze Fisiche e Scienze della Terra,\\viale F. Stagno d'Alcontres 31, 98166 Messina, Italy}
\date{\today}

\begin{abstract}
We study a system of penetrable bosons embedded in a spherical surface. Under the assumption of weak interaction between the particles, the ground state of the system is, to a good approximation, a pure condensate. We employ thermodynamic arguments to investigate, within a variational ansatz for the single-particle state, the crossover between distinct finite-size ``phases'' in the parameter space spanned by the sphere radius and the chemical potential. In particular, for radii up to a few interaction ranges we examine the stability of the fluid phase with respect to a number of crystal-like arrangements having the symmetry of a regular or semi-regular polyhedron. We find that, while quantum fluctuations keep the system fluid at low density, upon compression it eventually becomes inhomogeneous, i.e., particles gather together in clusters. As the radius increases, the nature of the high-density aggregate varies and we observe a sequence of transitions between different cluster phases (``solids''), whose underlying rationale is to maximize the coordination number of clusters, while ensuring at the same time the proper distance between each neighboring pair. We argue that, at least within our mean-field description, every cluster phase is supersolid.
\end{abstract}
\maketitle

\section{Introduction}

In the last few decades, thanks to the continued advance in the preparation and manipulation of ultracold atomic gases, the investigation of quantum correlation effects has reached a level of sophistication which would have been simply unimaginable before~\cite{Dalfovo,Leggett1,Bloch}. By confining atoms in optical and magnetic traps also the properties of low-dimensional quantum systems can be analyzed in detail, making it possible to test theoretical predictions and approximations (see, e.g., \cite{Kinoshita}). Even the range and strength of interatomic forces can be tuned to a certain extent~\cite{Buechler} (e.g., by the technique of Rydberg dressing), which has ultimately allowed to realize atomic systems characterized by an effective isotropic repulsion saturating to a {\em finite value} for zero separation~\cite{Henkel,Pupillo}.

Finite-strength interactions are frequently encountered in the classical realm as models for polymer coils or dendrimers dispersed in a good solvent (see, e.g., Refs.~\cite{Louis,Mladek}). The phase behavior of such fluids can be very rich, featuring in equilibrium any sort of mesoscopic structures (like clusters, micelles, and filaments --- see \cite{Rendiconti} and references cited therein). For purely repulsive particles, a distinctly universal behavior emerges at low temperature~\cite{Likos}, where, depending on the shape of the potential, the high-density phase is either fluid or cluster solid. The latter phase can be described as a crystalline system with multiply-occupied cells, each hosting on average the same number of particles (see examples in Refs.~\cite{Zhang,Prestipino1,Prestipino2}). Cluster crystals are characterized by a marked single-particle diffusion~\cite{Moreno}, which keeps the interstitial density at a non-zero value in equilibrium. It is clear that clustering, as a self-assembly phenomenon, can only occur when the formation of bunches of fully overlapping particles is energetically preferred over diffuse partial overlap~\cite{Mladek}.

In recent years, various quantum models of softly-repulsive bosons have been considered, whose phase diagram was worked out at zero temperature ($T=0$) both in mean field (MF)~\cite{Henkel2,Ancilotto,Kunimi,Prestipino3,Prestipino4} and by Monte Carlo (MC) simulation~\cite{Saccani,Cinti1,Cinti2,Macri}. In these systems the fluid-to-solid transition is the necessary outcome of the softening of roton-like modes in the fluid. The mechanism promoting quantum crystallization at $T=0$ is different from the freezing of hard-core fluids at high $T$, which is typically an entropy-driven (rather than an energy-driven) phase transition. Moreover, quantum cluster crystals may be supersolid, a feature which lacks a classical analog. Supersolidity has to do with an anomalous decrease of rotational inertia~\cite{Leggett2,Sepulveda,Kuklov,Boninsegni}, as if a fraction of the system remains stationary when the crystal is set into slow rotation around an axis.
 
In the present study we explore by MF theory the low-temperature physics of penetrable bosons in a setting which apparently has not been considered so far, i.e., that provided by confinement to a spherical surface. The systems which more closely resemble our model system are ultracold dilute gases trapped in a thin spherical shell, which have been the subject of a few experimental studies~\cite{Zobay,Garraway}. Other examples of real systems bearing some similarity to our model are multi-electron bubbles in liquid helium~\cite{Tempere}, arrangements of protein subunits on spherical viruses~\cite{Zandi}, and colloidal particles in colloidosomes~\cite{Fantoni}. In such systems the interparticle forces depend on the Euclidean distance rather than on the arc-length distance, which is the intrinsic metric for particles embedded in the surface of a sphere. However, this difference is immaterial as long as in our theory the pair potential is expressed in terms of the angular separation between the particles (see Section II). Spherical boundary conditions have often been used in numerical experiments~\cite{Prestipino5,Prestipino6,Prestipino7,Guerra,Post,Vest,Bozic} as a means to discourage crystalline ordering at high density (as well known, triangular order is frustrated on a sphere). In practice, the sphere curvature imposes a distinct excess of fivefold disclinations over sevenfold ones, which considerably complicates the search for optimal packings, even for small radii. Very recently, Franzini {\em et al.}~\cite{Franzini} have studied by density-functional theory a system of classical particles interacting through a generalized-exponential repulsion (GEM-4), finding a rich catalog of cluster phases as a function of the sphere radius $R$.

It is reasonable that, as the spherical surface gets more and more filled with particles, it will be found more convenient also for a quantum system of penetrable disks to clusterize, thus becoming solid-like inhomogeneous; moreover, as for a classical system, it is likely that numerous aggregates will compete for stability as a function of $R$. The most symmetric ones, i.e., those sharing the symmetries of a regular or semi-regular circumscribable polyhedron, are natural candidates for the high-density phases. There is a limiting case where the theoretical analysis of the quantum system at $T=0$ is greatly simplified, that is weak interparticle repulsion. Then, MF theory becomes an effective method, as practically demonstrated for a specific instance of soft-core bosons by the ``exact'' phase diagram reported in \cite{Cinti2}. As already made in \cite{Prestipino3,Prestipino4}, we further simplify our treatment using an educated guess of the condensate wave function, to be optimized by the variational method. By taking advantage of a well-established theoretical framework, we aim to gain insight into the self-organization principles underlying structure selection in a quantum many-body system characterized by a wealth of possible ground states.

The outline of the paper is as follows. In Sec.\,II we introduce the model and outline the variational MF theory employed to study its ground-state behavior. To give a flavour of the results obtained, in Sec.\,III we work out analytically a simpler exercise, which is nonetheless capable to predict the onset of clusters at high density in a specific range of $R$ values. Afterwards, in Sec.\,IV we present the full phase diagram of the system as a function of $R$ and chemical potential. In Sec.\,V we devote special attention to the issue of supersolidity of the spherical cluster phases. We show that, within our theory, all such phases are indeed supersolid. Concluding remarks are postponed to Sec.\,VI.

\section{Model and theory}
\setcounter{equation}{0}
\renewcommand{\theequation}{2.\arabic{equation}}

We investigate a system of $N$ identical spinless bosons, living on a sphere of radius $R$ and interacting with each other via a {\em bounded} potential $v(s)$, function of the arc-length distance $s$. A paradigmatic case of finite repulsion is the penetrable-sphere model (PSM) potential, $v(s)=\epsilon \vartheta(\sigma-s)$, $\vartheta$ being the Heaviside step function (PSM bosons will be our favourite case study later). It is convenient to introduce another parametrization of the potential, written in terms of the scalar product between the unit vectors $\hat{\bf r}$ and $\hat{\bf r}'$ identifying the positions on the sphere of the interacting pair. Using $\hat{\bf r}\cdot\hat{\bf r}'=\cos(s/R)$, we define:
\be
u(x)=v(R\arccos x)\,\,\,\,\,\,{\rm or}\,\,\,\,\,\,v(s)=u\left(\cos\frac{s}{R}\right)\,.
\label{eq2-1}
\ee
If the interaction potential were given in terms of the 3D Euclidean distance $r$, the definition of $u$ in (\ref{eq2-1}) would be modified into $u(x)=v(2R\sqrt{(1-x)/2})$, but no change will occur in the subsequent analysis.

It is not a priori obvious how to quantize a system of particles living in a curved space. Canonical quantization rules are inconsistent and a way out is to quantize angular momentum directly --- see this point thoroughly discussed in \cite{Kleinert}. In case of a free particle on a sphere, this entails taking the Hamiltonian (kinetic energy) to be $L^2/(2mR^2)=-\hbar^2/(2m)\nabla^2$, where $m$ is the particle mass and $\nabla^2$ is the Laplace-Beltrami operator on the sphere (see Eq.\,(\ref{eq2-4}) below). This same approach was followed by many authors~\cite{Ezra,Seidl,Loos,Yang}.

In the MF (Hartree) approximation, particles are treated as they were independent of each other and the $N$-boson ground state is therefore a pure condensate:
\be
\Psi(\Omega_1,\ldots,\Omega_N)=\prod_{i=1}^N\psi(\Omega_i)\,,
\label{eq2-2}
\ee
where $\Omega_i=(\theta_i,\phi_i)$ are the angular coordinates of the $i$-th particle, i.e., the spherical variables specifying its 3D position ${\bf r}_i=R\hat{\bf r}_i$ (in the following, $\Omega$ and $\hat{\bf r}$ are used interchangeably as argument of $\psi$). The best choice of single-particle wave function is that minimizing the expectation value of the Hamiltonian in $\Psi$, which corresponds to a normalized function obeying the (time-independent) Gross-Pitaevskii (GP) equation~\cite{Gross1,Pitaevskii1,Gross2} (see Appendix A):
\be
-\frac{\hbar^2}{2m}\nabla^2\psi+(N-1)\int{\rm d}^2\Omega'\,|\psi(\Omega')|^2u(\hat{\bf r}\cdot\hat{\bf r}')\psi(\Omega)=\lambda\psi(\Omega)\,,
\label{eq2-3}
\ee
where
\be
\nabla^2=\frac{1}{R^2}\left\{\frac{1}{\sin\theta}\frac{\partial}{\partial\theta}\left(\sin\theta\frac{\partial}{\partial\theta}\right)+\frac{1}{\sin^2\theta}\frac{\partial^2}{\partial\phi^2}\right\}
\label{eq2-4}
\ee
is the spherical Laplacian and
\be
\lambda=-\frac{\hbar^2}{2m}\int{\rm d}^2\Omega\,\psi^*\nabla^2\psi+(N-1)\int{\rm d}^2\Omega\,{\rm d}^2\Omega'\,|\psi(\Omega)|^2u(\hat{\bf r}\cdot\hat{\bf r}')|\psi(\Omega')|^2\,.
\label{eq2-5}
\ee
The value of $\lambda$ is consistent with $\psi$ being a solution to (\ref{eq2-3}): indeed, multiplication of both sides of (\ref{eq2-3}) by $\psi^*$ and subsequent integration over angles immediately leads to (\ref{eq2-5}) under the assumption of unit norm for $\psi$. In Appendix A, we discuss problems related with the numerical solution of Eq.\,(\ref{eq2-4}). We argue that a faster and physically more grounded method, which has proved effective in our exploration of the ground state of the planar system, is to minimize the following MF energy functional (kinetic energy per particle plus potential energy per particle) via the optimization of a parametric wave function:
\be
\label{eq2-6}
{\cal E}[\psi]=-\frac{\hbar^2}{2m}\int{\rm d}^2\Omega\,\psi^*\nabla^2\psi+\frac{N-1}{2}\int{\rm d}^2\Omega\,{\rm d}^2\Omega'\,|\psi(\Omega)|^2u(\hat{\bf r}\cdot\hat{\bf r}')|\psi(\Omega')|^2\,.
\ee
For a short-range potential, the ground-state energy in the planar limit $R\gg\sigma$ is only controlled by the dimensionless quantity $\rho\sigma^2\epsilon/e_0$ (where $e_0=\hbar^2/(m\sigma^2)$), which we hereafter refer to as the ``density''. When $R$ is finite the sphere radius is an additional control parameter, i.e., the properties of the system depend separately on $R$ and $\rho$. In the following, we take $\sigma$ and $e_0$ as units of length and energy, respectively.

We now provide a more explicit expression of ${\cal E}[\psi]$ that applies for any normalized wave function written as an expansion in spherical harmonics:
\be
\psi(\Omega)=\sum_{l=0}^\infty\sum_{m=-l}^lc_{lm}Y_l^m(\Omega)\,\,\,\,\,\,{\rm with}\,\,\,\,\,\,\sum_{lm}|c_{lm}|^2=1
\label{eq2-7}
\ee
(if $\psi(\Omega)$ is real then $c_{l,-m}=(-1)^mc_{lm}^*$). Computing the kinetic energy is straightforward; since $R^2\nabla^2Y_l^m=-l(l+1)Y_l^m$, we readily find:
\be
{\cal E}_{\rm kin}=\frac{\hbar^2}{2mR^2}\sum_{lm}l(l+1)|c_{lm}|^2\,.
\label{eq2-8}
\ee
As to the potential energy ${\cal E}_{\rm pot}$ (second term in Eq.\,(\ref{eq2-6})), in Appendix B we derive the following result:
{\small
\ba
{\cal E}_{\rm pot}&=&\frac{N-1}{4}\sum_{l=0}^\infty(2l+1)\int_{-1}^1{\rm d}x\,u(x)P_l(x)
\nonumber \\
&\times&\sum_{m=-l}^l(-1)^m\sum_{l_1m_1,l_2m_2,l_3m_3,l_4m_4}(-1)^{m_2+m_4}\sqrt{(2l_1+1)(2l_2+1)(2l_3+1)(2l_4+1)}
\nonumber \\
&\times&\left(\begin{array}{ccc}l & l_1 & l_2 \\0 & 0 & 0\end{array}\right)\left(\begin{array}{ccc}l & l_3 & l_4 \\0 & 0 & 0\end{array}\right)\left(\begin{array}{ccc}l & l_1 & l_2 \\m & m_1 & -m_2\end{array}\right)\left(\begin{array}{ccc}l & l_3 & l_4 \\-m & m_3 & -m_4\end{array}\right)c_{l_1m_1}c_{l_2m_2}^*c_{l_3m_3}c_{l_4m_4}^*\,,
\nonumber \\
\label{eq2-9}
\ea
}
where $P_l(x)$ are Legendre polynomials and the matrices are Wigner 3-j symbols.

The total energy per particle is the sum of (\ref{eq2-8}) and (\ref{eq2-9}). In practice, the $l$ sum must be truncated, i.e., $l\le l_{\rm max}$, where $l_{\rm max}$ is chosen in accordance with the spatial resolution adopted for the description (see this point discussed, e.g., in Ref.~\cite{Prestipino8}). To check consistency, let us consider the homogeneous fluid, corresponding to $c_{lm}=\delta_{l0}\delta_{m0}$ or $\psi=Y_0^0=1/\sqrt{4\pi}$. It then follows from Eqs.\,(\ref{eq2-8}) and (\ref{eq2-9}), as well as directly from Eq.\,(\ref{eq2-6}), that 
\be
{\cal E}=\frac{N-1}{4}\int_{-1}^1{\rm d}x\,u(x)=\frac{N-1}{4}\int_{-1}^1{\rm d}x\,v(R\arccos x)=\frac{N-1}{8\pi}\int{\rm d}^2\Omega\,v(R\theta)\longrightarrow\frac{\rho}{2}\widetilde{v}(0)\,,
\label{eq2-10}
\ee
where the last step follows in the planar limit, that is for $N\rightarrow\infty,\,R\rightarrow +\infty$, and $N/(4\pi R^2)\rightarrow\rho$. As expected, the limiting value of ${\cal E}$ is nothing but the specific energy of the planar fluid\,\cite{Prestipino3}.

\section{Cluster formation at high density: Proof of concept}
\setcounter{equation}{0}
\renewcommand{\theequation}{3.\arabic{equation}}

As argued in the Introduction, in a spherical quantum system the most stable $T=0$ configuration would not necessarily be fluid. Depending on the radius $R$, other phases may be expected to arise as ground states in a system of softly-repulsive particles at high density. In particular, we guess a primary role for cluster-crystal-like arrangements having the symmetry of a regular (i.e., Platonic) or semi-regular (i.e., Archimedean or circumscribable Catalan) polyhedron. It is easy to conjecture that the stable phase at high density would crucially depend on the value of $R$, since the latter determines the edge length $\ell$ of the inscribed polyhedron and consequently also the geodesic distance between two neighboring clusters. Considering that for the PSM interaction the edge of the triangular-crystal lattice is about $1.51\sigma$ at melting~\cite{Prestipino3}, we expect that the structure of the high-density phase will be found among those polyhedra having $\ell\lesssim 1.51\sigma$. For instance, since the sphere circumscribing the regular icosahedron has a radius of
\be
R=\frac{\ell}{4}\sqrt{10+2\sqrt{5}}\,,
\label{eq3-1}
\ee
an icosahedral cluster phase is most likely to occur for $R\approx 1.4\sigma$.

A real, one-parameter form of $\psi$ adequate to represent the pattern expected at large $\rho$ on the sphere is a sum of Gaussians centered at the vertices ${\bf R}_k$ $(k=1,\ldots,n)$ of the inscribed polyhedron:
\be
\psi(\hat{\bf r})=C_\alpha\sum_{k=1}^{n}\exp\left\{-\alpha\left(\frac{R\hat{\bf r}-{\bf R}_k}{\ell}\right)^2\right\}=C_\alpha\sum_{k=1}^{n}\exp\left\{-\frac{2R^2}{\ell^2}\alpha(1-\hat{\bf r}\cdot\hat{\bf R}_k)\right\}\,,
\label{eq3-2}
\ee
where $\alpha$ is a variational parameter and $C_\alpha$ is a (real) normalization constant (notice that $R/\ell$ is a pure number, specific of the given polyhedron). The fluid ground state ($\psi=1/\sqrt{4\pi}$) is recovered from (\ref{eq3-2}) as a limiting case, i.e., for $\alpha=0$. In a solid-like system, $\alpha>0$ represents the inverse square width of the local-density peaks. We point out that a real $\psi$ is not a limitation whatsoever; indeed, we show in Appendix C that the true single-particle wave function of minimum energy is, as already known from three dimensions~\cite{Pomeau}, a real function.

We now describe the method followed to draw the ``melting line'' of a given cluster phase as a function of $R$ for $T=0$. To this purpose we employ a thermodynamic framework, with the idea that when a crossover (a rounded phase transition) occurs at fixed $R$ from one ground state to the other the number of particles is very large (we shall see {\em a posteriori} that this is always a safe assumption). In this respect, the chemical potential $\mu$ is a more meaningful control parameter than the pressure $P$ since the surface area is fixed. In brief, we first determine the energy per unit particle $e$ as a function of $\rho$, taking $N=4\pi R^2\rho$ in Eq.\,(\ref{eq2-6}). Called $\overline{\alpha}(\rho)$ the point of absolute minimum of ${\cal E}([\psi(\alpha)];\rho)$ for the fixed $\rho$, we have $e(\rho)={\cal E}([\psi(\overline{\alpha}(\rho)];\rho)$. Once the energy has been computed, the transition point $\mu_c(R)$ is where the fluid and the solid have the same grand potential per unit area (i.e., where the minimum of $\rho(e(\rho)-\mu)$ is the same for both phases). In the fluid phase, where Eq.\,(\ref{eq2-10}) holds, the relation between $\mu$ and $\rho$ is thus found to be $\mu=(2+(N-1)^{-1}){\cal E}(\rho)\simeq 2{\cal E}(\rho)$.

In order to compute ${\cal E}[\psi]$, two different roads can be followed: either we evaluate Eq.\,(\ref{eq2-6}) numerically, or we attempt an estimate of the Fourier coefficient
\be
c_{lm}=\int{\rm d}^2\Omega\,Y_l^{m*}(\Omega)\psi(\Omega)
\label{eq3-3}
\ee
for all $l\le l_{\rm max}$, and then use Eqs.\,(\ref{eq2-8}) and (\ref{eq2-9}). Indeed, in the following we will pursue both routes; however, before that we show the feasibility of our approach by providing a fully analytic demonstration of clusterization in a system of spherical bosons at $T=0$.

For our proof we make use of a variational wave function simpler than (\ref{eq3-2}), but still endowed with the symmetries of the high-density phase we aim to describe (we will focus on the {\em icosahedral} cluster phase). In this regard, it is useful to recall an important paper by Zheng and Doerschuk\,\cite{Zheng} where they explain how to construct a basis in the subspace of square-integrable $\Omega$ functions that are invariant under every rotation of the icosahedral group (see related comments on this subject at \cite{Baez,Egan}). These basis functions, denoted $T_l^n(\Omega)$ and dubbed {\em icosahedral harmonics}, are real and orthonormal, and given by
\be
T_l^n(\Omega)=\sum_{m=-l}^lb_{nlm}Y_l^m(\Omega)\,.
\label{eq3-4}
\ee
For fixed $l$ there are $N_l$ icosahedral harmonics $T_l^n$ ($n=0,\ldots,N_l-1$) that are linear combinations of the $Y_l^m$ for $m=-l,\ldots,l$ (hence $N_l\le 2l+1$). Zheng and Doerschuk have derived recursive formulae for the coefficients $b_{nlm}$, for arbitrary $n,l,m$, including the cases where $N_l>1$ (which only occurs for $l\ge 30$). In particular, the first three icosahedral harmonics turn out to be:
\ba
T_0^0&=&Y_0^0\,;\,\,\,T_6^0=\frac{\sqrt{7}}{5}Y_6^{-5}+\frac{\sqrt{11}}{5}Y_6^0-\frac{\sqrt{7}}{5}Y_6^5\,;
\nonumber \\
T_{10}^0&=&\frac{\sqrt{187}}{25\sqrt{3}}Y_{10}^{-10}-\frac{\sqrt{209}}{25}Y_{10}^{-5}+\frac{\sqrt{247}}{25\sqrt{3}}Y_{10}^0+\frac{\sqrt{209}}{25}Y_{10}^5+\frac{\sqrt{187}}{25\sqrt{3}}Y_{10}^{10}\,.
\label{eq3-5}
\ea
Of particular interest to us is the function $T_6^0$, which has twelve maxima of same height centered at the vertices of a regular icosahedron (see Fig.\,1). Hence, our problem becomes one of determining whether, at sufficiently high density, the energy of
\be
\psi=C_\beta(T_0^0+\beta T_6^0)\,\,\,\,\,\,\,\,\,\,\left({\rm with}\,\,\,C_\beta=\frac{1}{\sqrt{1+\beta^2}}\right)
\label{eq3-6}
\ee
reaches its minimum for some $\beta>0$. In this circumstance, the fluid phase (represented by $T_0^0$) is doomed to transform upon compression into an icosahedral cluster phase. The only caveat is that $\beta<\beta_{\rm max}\simeq 0.7$ in (\ref{eq3-6}), if we want to exclude the appearance of spurious maxima in $\psi^2$.

%
%
\begin{figure}
\begin{center}
\includegraphics[width=8cm]{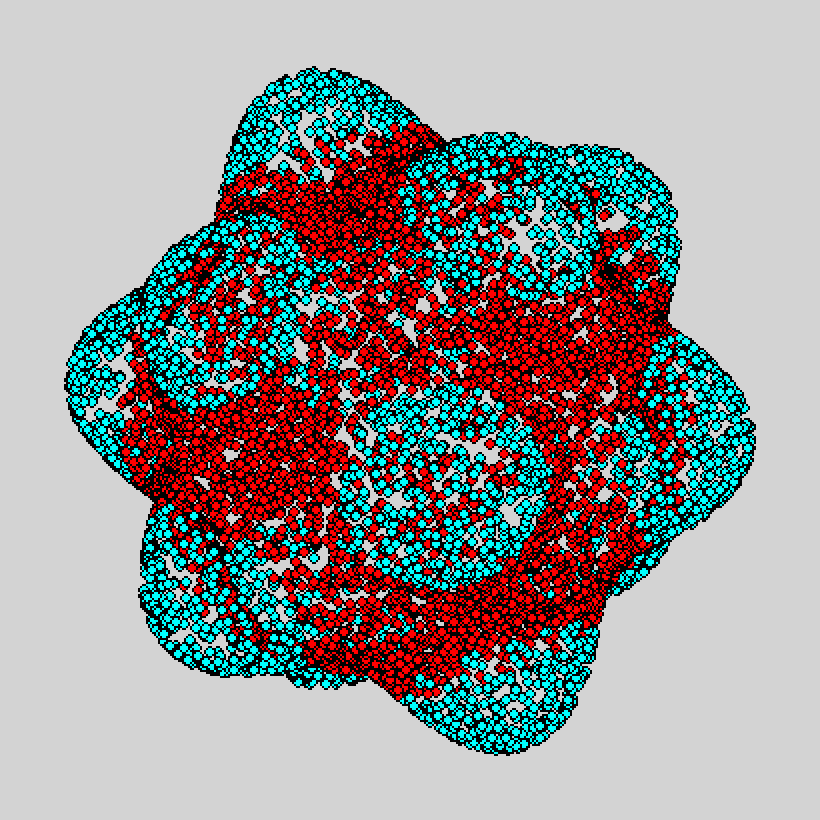}
\includegraphics[width=8cm]{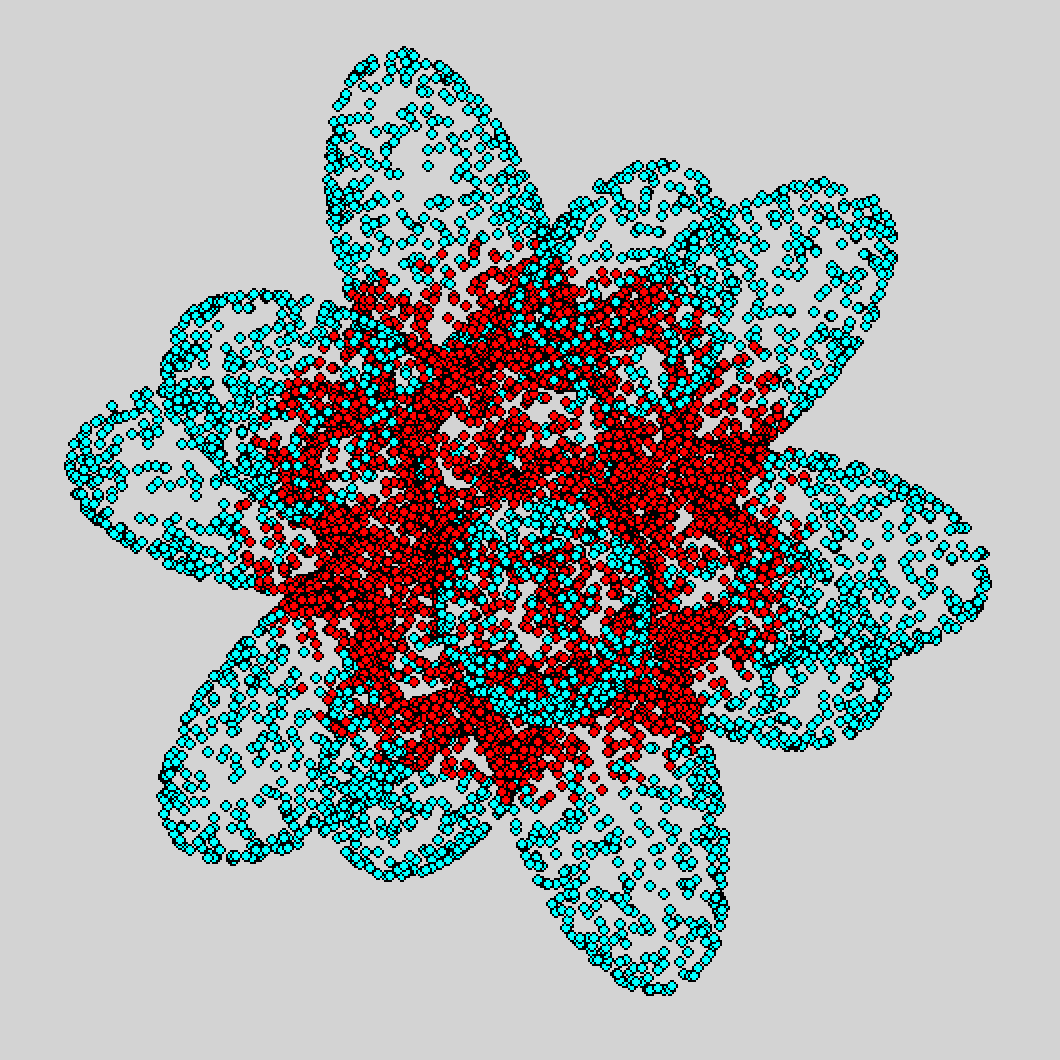}
\end{center}
\caption{Scatter plot of the surface ${\cal S}:r=T_0^0+\beta T_6^0(\theta,\phi)$, for $\beta=0.1$ (left) and 0.3 (right), which clearly reveals the icosahedral symmetry. As $\beta$ increases, clusters become more localized on the sphere. These plots were obtained by picking $10^4$ vectors $(x,y,z)$ at random over the unit sphere and multiplying each by the respective value of $T_0^0+\beta T_6^0$. The surface ${\cal S}$ is the envelope of the free ends of the vectors. The red (blue) points are those $(x,y,z)$ where $T_6^0>0$ ($T_6^0<0$, respectively).}
\label{fig1}
\end{figure}

While it is immediate to compute the kinetic energy per particle (cf. Eq.\,(\ref{eq2-8})), equal to
\be
{\cal E}_{\rm kin}=21\beta^2C_\beta^2\frac{\hbar^2}{mR^2}\,,
\label{eq3-7}
\ee
it is much harder to determine ${\cal E}_{\rm pot}$, given by the double integral
\be
\frac{N-1}{2}C_\beta^4\int{\rm d}^2\Omega\,{\rm d}^2\Omega'\left(T_0^0+\beta T_6^0(\hat{\bf r})\right)^2u(\hat{\bf r}\cdot\hat{\bf r}')\left(T_0^0+\beta T_6^0(\hat{\bf r}')\right)^2\,.
\label{eq3-8}
\ee
After expanding each square, Eq.\,(\ref{eq3-8}) becomes the sum of nine terms, not all distinct, most of which can be simply evaluated by using the orthonormality property of spherical harmonics:
\ba
&&\int{\rm d}^2\Omega\,{\rm d}^2\Omega'\,(T_0^0)^2u(\hat{\bf r}\cdot\hat{\bf r}')(T_0^0)^2=\frac{1}{2}\int_{-1}^1{\rm d}x\,u(x)\,;
\nonumber \\
&&\int{\rm d}^2\Omega\,{\rm d}^2\Omega'\,2\beta T_0^0T_6^0(\hat{\bf r})u(\hat{\bf r}\cdot\hat{\bf r}')(T_0^0)^2=0\,;
\nonumber \\
&&\int{\rm d}^2\Omega\,{\rm d}^2\Omega'\,\beta^2(T_6^0(\hat{\bf r}))^2u(\hat{\bf r}\cdot\hat{\bf r}')(T_0^0)^2=\frac{\beta^2}{2}\int_{-1}^1{\rm d}x\,u(x)\,;
\nonumber \\
&&\int{\rm d}^2\Omega\,{\rm d}^2\Omega'\,2\beta T_0^0T_6^0(\hat{\bf r})u(\hat{\bf r}\cdot\hat{\bf r}')2\beta T_0^0T_6^0(\hat{\bf r}')=2\beta^2\int_{-1}^1{\rm d}x\,u(x)P_6(x)\,.
\label{eq3-9}
\ea
More cumbersome is the calculation of
\be
I_1=\int{\rm d}^2\Omega\,{\rm d}^2\Omega'\,\beta^2(T_6^0(\hat{\bf r}))^2u(\hat{\bf r}\cdot\hat{\bf r}')2\beta T_0^0T_6^0(\hat{\bf r}')
\label{eq3-10}
\ee
and
\be
I_2=\int{\rm d}^2\Omega\,{\rm d}^2\Omega'\,\beta^2(T_6^0(\hat{\bf r}))^2u(\hat{\bf r}\cdot\hat{\bf r}')\beta^2(T_6^0(\hat{\bf r}'))^2\,.
\label{eq3-11}
\ee
In the former case, we are required to compute 3-j symbols of the kind
\be
\left(\begin{array}{ccc}6 & 6 & 6 \\m & m_1 & -m_2\end{array}\right)
\label{eq3-12}
\ee
with $m,m_1,m_2=0,\pm 5$. The only non-zero symbols are those for which $m+m_1-m_2=0$, which can occur in one of seven ways. The end result is
\be
I_1=\frac{20\sqrt{11\cdot 13}}{17\cdot 19}\beta^3\int_{-1}^1{\rm d}x\,u(x)P_6(x)\,.
\label{eq3-13}
\ee
As for $I_2$, we directly start from Eq.\,(\ref{eq2-9}) with $\psi=\beta T_6^0$. We now need to compute the following 3-j symbols:
\be
\left(\begin{array}{ccc}l & 6 & 6 \\0 & 0 & 0\end{array}\right)\,,\,\,\,\left(\begin{array}{ccc}l & 6 & 6 \\m & m_1 & -m_2\end{array}\right)\,,\,\,\,\left(\begin{array}{ccc}l & 6 & 6 \\-m & m_3 & -m_4\end{array}\right)\,.
\label{eq3-14}
\ee
The first symbol is non-zero exclusively for $l$ even and not larger than 12. For each allowed $l$, the second symbol in Eq.\,(\ref{eq3-14}) is non-zero in at most nine cases (three cases for $l=0,2,4$, seven for $l=6,8$, and nine for $l=10,12$):
\ba
&&m=m_1=m_2=0\,;
\nonumber \\
&&m=0,m_1=m_2=5\,;
\nonumber \\
&&m=0,m_1=m_2=-5\,;
\nonumber \\
&&m=5,m_1=-5,m_2=0\,;
\nonumber \\
&&m=5,m_1=0,m_2=5\,;
\nonumber \\
&&m=-5,m_1=5,m_2=0\,;
\nonumber \\
&&m=-5,m_1=0,m_2=-5\,;
\nonumber \\
&&m=10,m_1=m_2=-5\,;
\nonumber \\
&&m=-10,m_1=m_2=5\,,
\label{eq3-15}
\ea
and similar considerations apply for the third 3-j symbol.

After a lengthy series of steps we are eventually led to:
\ba
{\cal E}_{\rm pot}&=&\frac{N-1}{2}C_\beta^4\left\{(1+\beta^2)^2E_0+\beta^4(E_2+E_4+E_6+E_8+E_{10}+E_{12})\right.
\nonumber \\
&+&\left.\left(2\beta^2+\frac{40\sqrt{11\cdot 13}}{17\cdot 19}\beta^3\right)\int_{-1}^1{\rm d}x\,u(x)P_6(x)\right\}
\label{eq3-16}
\ea
with
\ba
E_0&=&\frac{1}{2}\int_{-1}^1{\rm d}x\,u(x)\,;\,\,\,E_2=E_4=0\,;\,\,\,E_6=\frac{2^3\cdot 5^2\cdot 11\cdot 13}{17^2\cdot 19^2}\int_{-1}^1{\rm d}x\,u(x)P_6(x)\,;\,\,\,E_8=0\,;
\nonumber \\
E_{10}&=&\frac{2\cdot 3^4\cdot 7^3\cdot 13}{17^2\cdot 19\cdot 23^2}\int_{-1}^1{\rm d}x\,u(x)P_{10}(x)\,;\,\,\,E_{12}=\frac{2^3\cdot 3^2\cdot 7^3\cdot 11^2}{5\cdot 17\cdot 19^2\cdot 23^2}\int_{-1}^1{\rm d}x\,u(x)P_{12}(x)\,.
\nonumber \\
\label{eq3-17}
\ea
We underline that the individual integrals contributing to the total energy have been successfully checked, for a few $\beta$ values, against MC integration (see Sec.\,IV).

For the PSM potential,
\be
u(x)=\left\{
\begin{array}{cc}
\epsilon, & R\arccos x\le\sigma \\
0, & R\arccos x>\sigma
\end{array}
\right.=\left\{
\begin{array}{cc}
0, & -1\le x<\cos(\sigma/R) \\
\epsilon, & \cos(\sigma/R)\le x \le 1
\end{array}
\right.\,,
\label{eq3-18}
\ee
it holds:
\be
\int_{-1}^1{\rm d}x\,u(x)P_l(x)=\epsilon\int_{\cos(\sigma/R)}^1{\rm d}x\,P_l(x)\,.
\label{eq3-19}
\ee
Using the property (valid for any $l>0$):
\be
(2l+1)P_l(x)=\frac{\rm d}{{\rm d}x}\left[P_{l+1}(x)-P_{l-1}(x)\right]\,,
\label{eq3-20}
\ee
we easily obtain:
\be
\int_{-1}^1{\rm d}x\,u(x)P_l(x)=\frac{\epsilon}{2l+1}\left[P_{l-1}\left(\cos\frac{\sigma}{R}\right)-P_{l+1}\left(\cos\frac{\sigma}{R}\right)\right]\,.
\label{eq3-21}
\ee
In other words, the PSM energy for $\psi=C_\beta(T_0^0+\beta T_6^0)$ admits an {\em explicit expression in closed form}.

%
%
\begin{figure}
\begin{center}
\includegraphics[width=10cm]{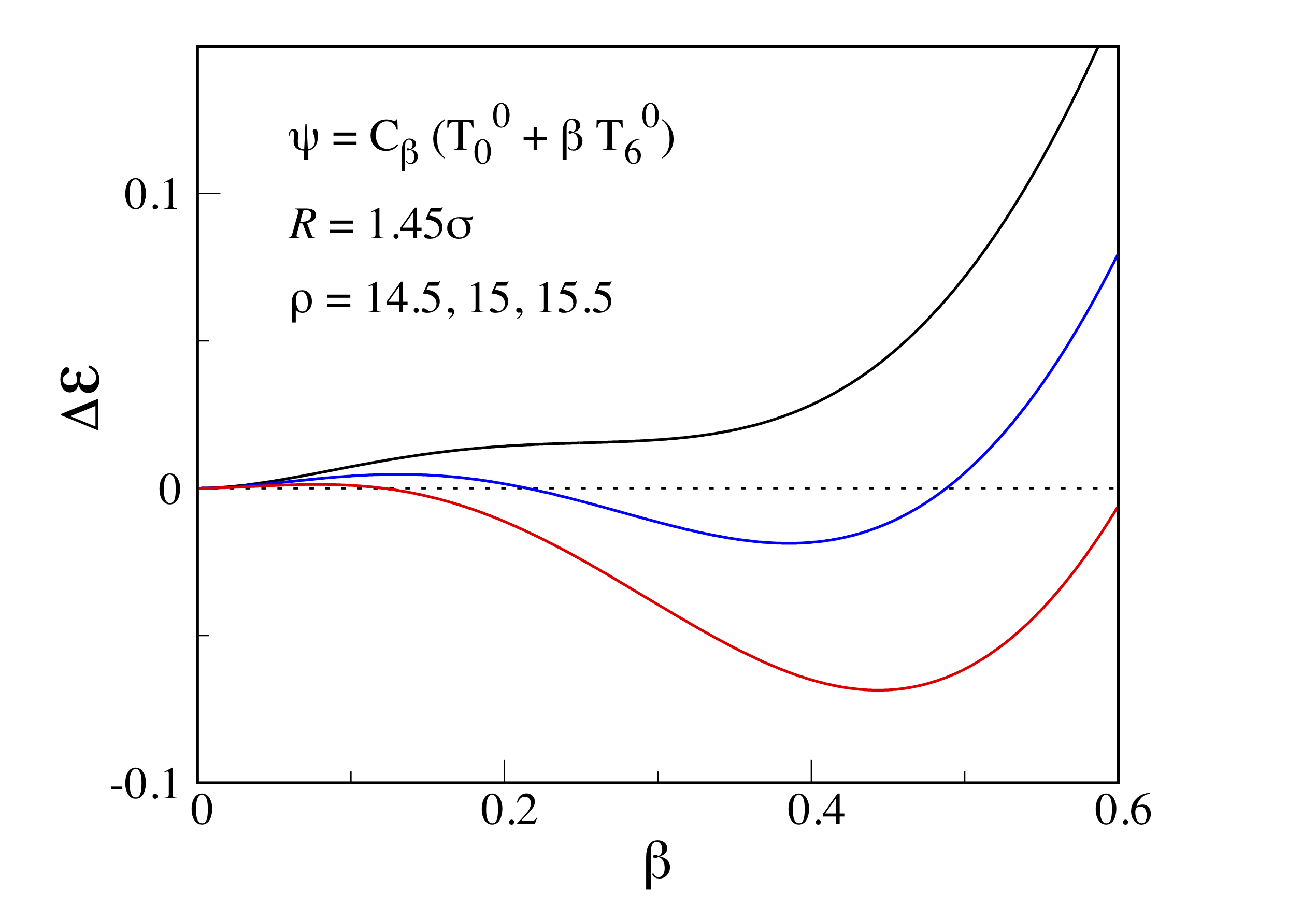}
\end{center}
\caption{Excess energy (units of $e_0$) of the PSM ``solid'' relative to the fluid, plotted as a function of $\beta$ for $R=1.45\sigma$ and three densities $\rho$ (from now on reported in reduced units $\sigma^{-2}e_0/\epsilon$): from top to bottom, $\rho=14.5$ (black), 15 (blue), and 15.5 (red).}
\label{fig2}
\end{figure}

We show in Fig.\,2 some data obtained from Eqs.\,(\ref{eq3-7}), (\ref{eq3-16}), (\ref{eq3-17}), and (\ref{eq3-21}). The plotted quantity is the excess energy $\Delta{\cal E}={\cal E}[\psi(\beta)]-{\cal E}_f$ with ${\cal E}_f=(N-1)E_0/2$, namely the energy of the solid relative to the fluid, which has been computed for $R=1.45\sigma$ and three distinct values of $\rho$. As $\rho$ increases, the solid energy falls eventually below the fluid energy, implying a transition from the fluid to the icosahedral phase upon varying $\mu$ at fixed $R$ (in this case, each cluster hosts roughly 30 particles at melting). This transition has the nature of a first-order phase change, accompanied by metastability of both phases beyond the transition point. It is worth noting the resemblance of this phenomenon to the onset of icosahedral ordering of {\em disordered aggregates of disclinations} in a system of hard calottes on a sphere~\cite{Prestipino5,Prestipino6,Guerra}. Even in that case a geometric pattern emerges when tuning a control parameter (the density), although its mechanism is purely entropic rather than energy-promoted as in the present case.

\section{Results}
\setcounter{equation}{0}
\renewcommand{\thesubsection}{\arabic{subsection}}
\renewcommand{\theequation}{4.\arabic{equation}}

%
%
\begin{figure}
\begin{center}
\includegraphics[width=16cm]{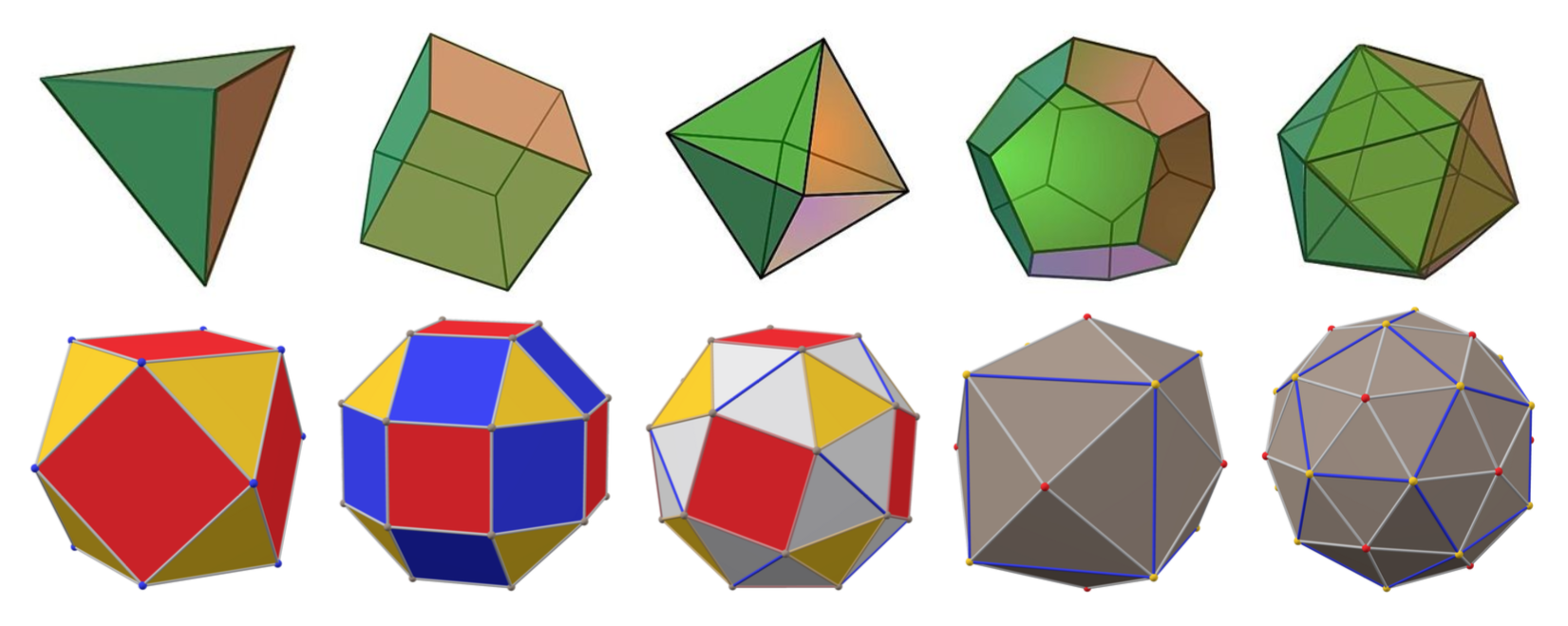}
\end{center}
\caption{The ten circumscribable polyhedra considered in this work. Each of them provides the underlying skeleton of a possible $T=0$ phase in a system of spherical bosons (i.e., clusters are centered at the vertices of the polyhedron). First row: the five Platonic solids (from left to right: tetrahedron, cube, octahedron, dodecahedron, and icosahedron). Second row: three Archimedean solids (from left to right: cuboctahedron, rombicuboctahedron, and snub cube) and two Catalan solids (tetrakis hexahedron and pentakis dodecahedron).}
\label{fig3}
\end{figure}

%
%
\begin{table}
\caption{Ratio between the edge length $\ell$ and the circumscribed radius $R$ for the polyhedra depicted in Fig.\,3.  For the snub-cube case, $t=\left(1+\sqrt[3]{19-3\sqrt{33}}+\sqrt[3]{19+3\sqrt{33}}\right)/3$ and $\beta=\sqrt[3]{26+6\sqrt{33}}$. In the last two lines, the quoted $\ell/R$ refers to the biscribed form of the polyhedron, and $\ell$ is the short edge (the most numerous one).}
\begin{center}
\begin{tabular}{ccc}
\hline\hline
polyhedron & $\ell/R$ & $D=2R^2/\ell^2$\\
\hline\hline
tetrahedron & \,\,\,\,\,\,$2\sqrt{2/3}=1.632\ldots$\,\,\,\,\,\, & $3/4$\\
cube & \,\,\,\,\,\,$2\sqrt{3}/3=1.154\ldots$\,\,\,\,\,\, & $3/2$\\
octahedron & \,\,\,\,\,\,$\sqrt{2}=1.414\ldots$\,\,\,\,\,\, & $1$\\
dodecahedron & \,\,\,\,\,\,$4/(\sqrt{3}+\sqrt{15})=0.713\ldots$\,\,\,\,\,\, & $(9+3\sqrt{5})/4=3.927\ldots$\\
icosahedron & \,\,\,\,\,\,$4/\sqrt{10+2\sqrt{5}}=1.051\ldots$\,\,\,\,\,\, & $(5+\sqrt{5})/4=1.809\ldots$\\
\hline
cuboctahedron & \,\,\,\,\,\,$1$\,\,\,\,\,\, & $2$\\
rombicuboct. & \,\,\,\,\,\,$2/\sqrt{5+2\sqrt{2}}=0.714\ldots$\,\,\,\,\,\, & $(5+2\sqrt{2})/2=3.914\ldots$\\
snub cube & \,\,\,\,\,\,$\sqrt{2(2-8/\beta+\beta)/3}/\sqrt{t^2+t^{-2}+1}=0.744\ldots$\,\,\,\,\,\, &\\
\hline
tetrakis hex. & \,\,\,\,\,\,$\sqrt{6(3-\sqrt{3})}/3=0.919\ldots$\,\,\,\,\,\, & $(3+\sqrt{3})/2=2.366\ldots$\\
pentakis dod. & \,\,\,\,\,\,$\sqrt{30\left(15-\sqrt{15(5+2\sqrt{5})}\right)}/15=0.640\ldots$\,\,\,\,\,\, & $$\\
\hline\hline
\end{tabular}
\end{center}
\end{table}

The main value of the simplified variational calculation carried out in Sec.\,III is to make it evident that in a system of weakly-repulsive spherical bosons a sharp crossover occurs at $T=0$, as a function of $\mu$ and for $R\approx 1.4\sigma$, from the fluid to a cluster phase of icosahedral symmetry. We emphasize that the choice of the icosahedron as supporting frame for the clusters is just one possibility; in fact, as $R$ increasingly departs from $1.4\sigma$, other polyhedra will be better suited than the icosahedron to match the condition $\ell\lesssim 1.51\sigma$. In Table I we report the value of $X=\ell/R$ for ten different solids, depicted in Fig.\,3, which we have selected among tens of regular or semi-regular circumscribable polyhedra~\cite{VisualPolyhedra} as the reference structures that likely underlie the cluster phases for $R$ up to $\approx 2.5\sigma$; each such geometry would become relevant in a $R$ interval centered about $1.51\sigma/X$. Besides the five regular (Platonic) solids, we focus our attention on three Archimedean solids and two biscribed Catalan solids (we recall that Archimedean solids have regular faces --- not all of the same type --- meeting in identical vertices, while Catalan solids are dual to Archimedean solids and not all vertices are equivalent). In selecting the solids in Table I our criterion was to rule out all semi-regular polyhedra with too many vertices (more than 32) or too large faces (which would correspond to big surface ``holes'' devoid of particles). Clearly, we have no argument to exclude that other structures will also come into play (in fact, we have good reasons to think that some low-symmetry structures are actually relevant, see more below), but there is anyway no hope to identify all local minima in what is likely to be a rugged free-energy landscape.

To draw the $T=0$ phase diagram of the spherical PSM system by the variational method, we need to evaluate the energy per particle (\ref{eq2-6}) with $\psi$ given as in Eq.\,(\ref{eq3-2}). A viable method is to resort to MC integration. For any of the structures in Table I, the specific energy  $\cal E$ at fixed $R$ is immediately obtained for any $\rho$ once the kinetic energy per particle ${\cal E}_{\rm kin}$ and the potential energy per pair, $2{\cal E}_{\rm pot}/(N-1)$, are given. Using a standard algorithm to generate points $\hat{\bf r}_i$ distributed at random over the unit sphere~\cite{Krauth}, we can estimate the two above-cited energies by the following formulae:
\be
{\cal E}_{\rm kin}\simeq-\frac{\hbar^2}{2m}\,\frac{\sum_i\psi_i(\nabla^2\psi)_i}{\sum_i\psi_i^2}\,\,\,\,\,\,{\rm and}\,\,\,\,\,\,\frac{2{\cal E}_{\rm pot}}{N-1}\simeq\frac{\sum_{i,j}\psi_i^2u_{ij}\psi_j^2}{\sum_{i,j}\psi_i^2\psi_j^2}
\label{eq4-1}
\ee
with $u_{ij}=u(\hat{\bf r}_i\cdot\hat{\bf r}_j)\,,\psi_i=\sum_{k=1}^{n}\exp\left\{-D\alpha(1-\hat{\bf r}_i\cdot\hat{\bf R}_k)\right\}$, and
\be
(\nabla^2\psi)_i=-\frac{D\alpha}{R^2}\sum_{k=1}^{n}\left\{2\hat{\bf r}_i\cdot\hat{\bf R}_k-D\alpha\left[1-\left(\hat{\bf r}_i\cdot\hat{\bf R}_k\right)^2\right]\right\}\exp\left\{-D\alpha(1-\hat{\bf r}_i\cdot\hat{\bf R}_k)\right\}\,.
\label{eq4-2}
\ee
In the above expressions, $n$ denotes the number of vertices of the given polyhedron while $D$ is a shorthand for $2R^2/\ell^2$. To make the error on the kinetic-energy estimate comparable to that on the potential energy, the total number ${\cal N}_{\rm tot}$ of points $i$ in the first of Eqs.\,(\ref{eq4-1}) has been taken equal to the number of random pairs in the second (typically ${\cal N}_{\rm tot}\approx 10^{10}$).

Figure 4 gives an idea of the type of results obtained. These data are relative to a pure condensate with snub-cube symmetry and refer to $R=1.9\sigma$. On the left panel of Fig.\,4, we have plotted the excess energy $\Delta{\cal E}={\cal E}-{\cal E}_f$ as a function of $\alpha$ for a few densities; we see that, for $\rho\gtrsim 14$, a minimum develops for a non-zero $\alpha$, signaling the onset of an inhomogeneous phase at high density. On the right panel, the chemical potential has been adjusted so that the grand potential of the fluid (hence, its thermodynamic pressure) equals that of the cluster phase. This condition defines the snub-cube transition point $\mu_c$ for the given $R$, whereas the abscissae of the two equal minima are the coexistence densities, $\rho_f$ and $\rho_s$. For this case, the average number of particles per cluster at melting is $4\pi R^2\rho_s/24=30.38$.

%
%
\begin{figure}
\begin{center}
\includegraphics[width=8cm]{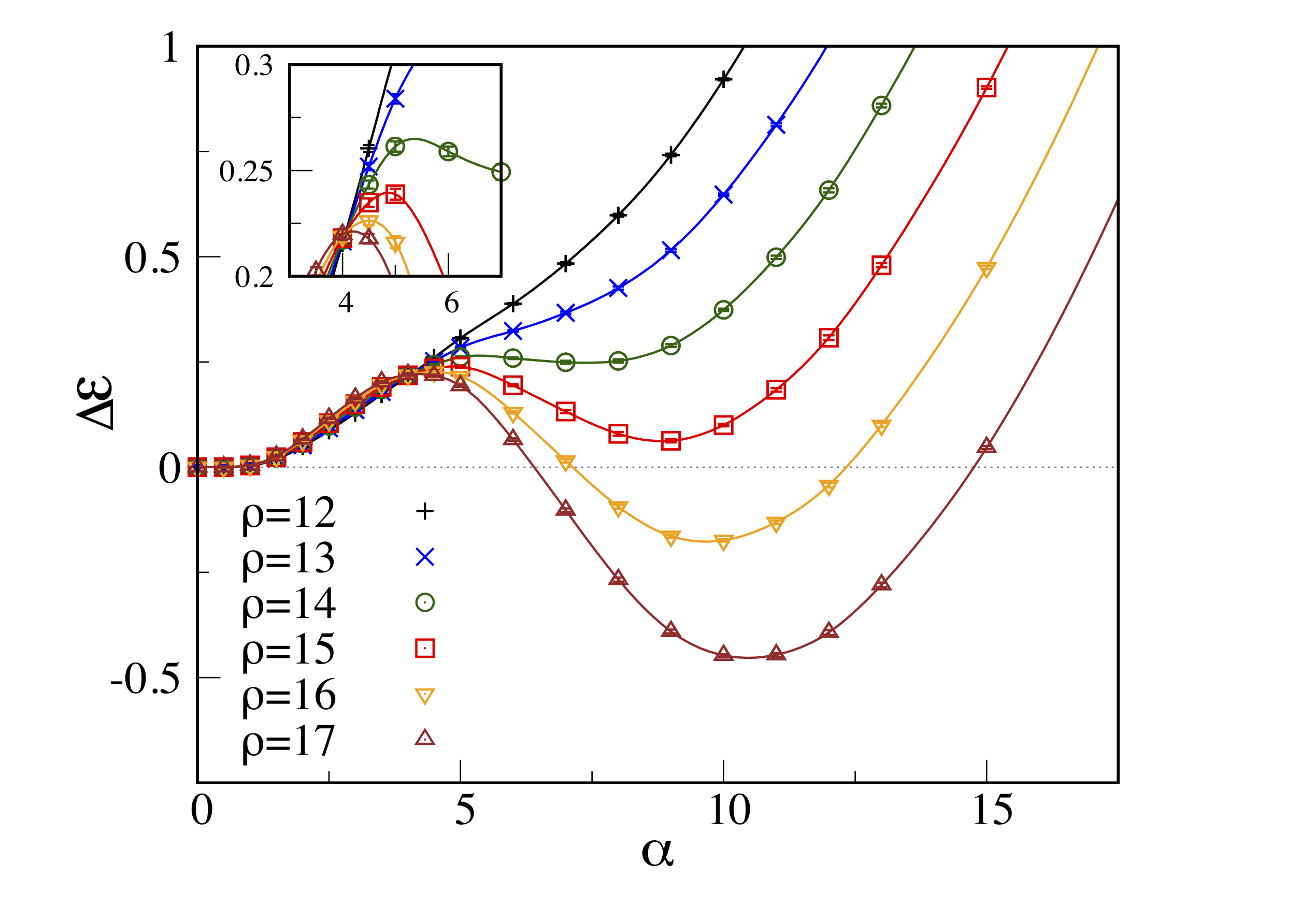}
\includegraphics[width=8cm]{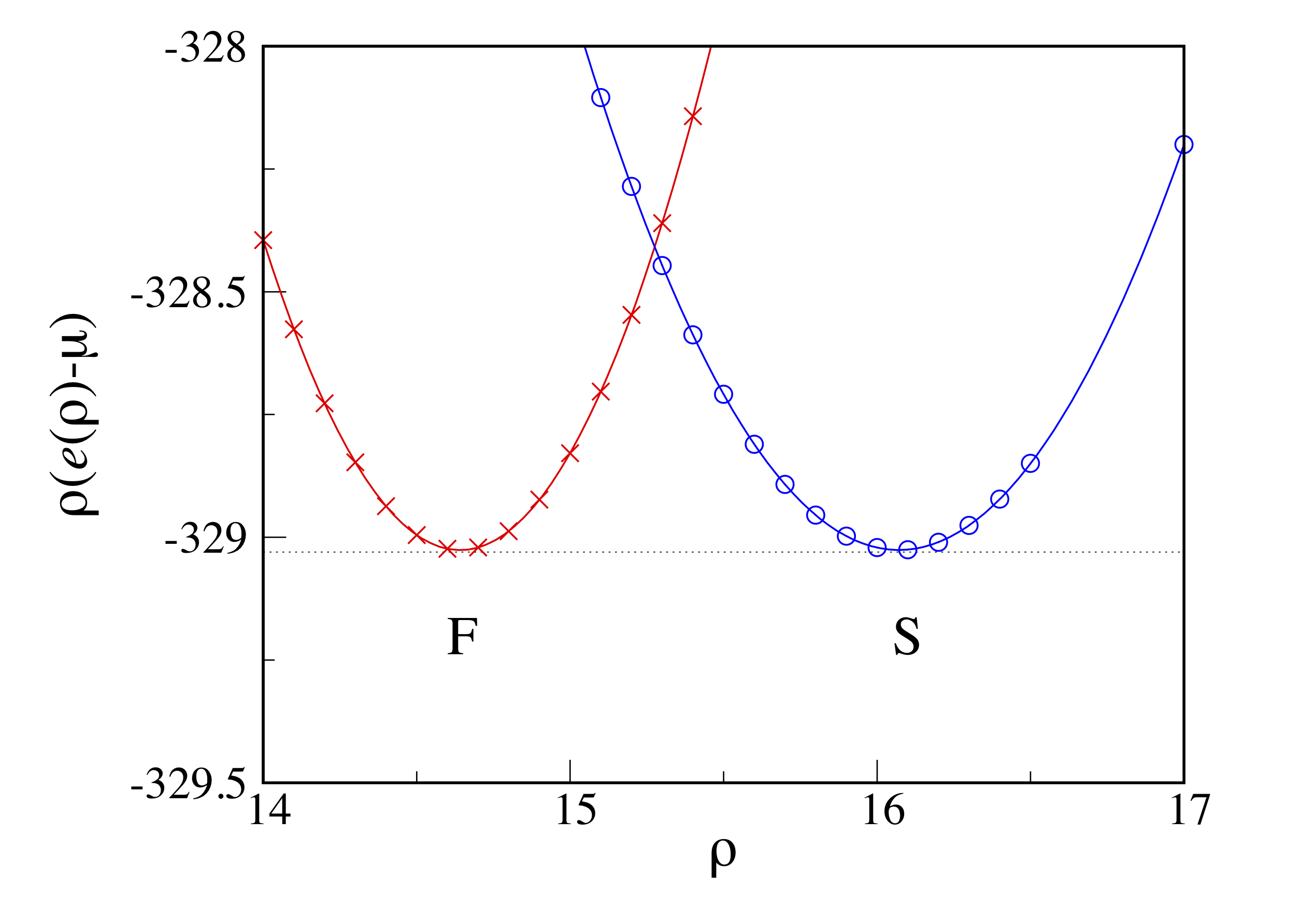}
\end{center}
\caption{PSM bosons on a sphere at $T=0$: an example of the determination of the phase-transition point (this case refers to an inhomogeneous phase having the symmetry of a snub cube, for $R=1.9\sigma$). Left: Excess energy (units of $e_0$) vs. $\alpha$ for a number of reduced densities in the range from 12 to 17 (each plotted point is the result of an average over 20 independent estimates of the energy given by Eq.\,(\ref{eq4-1})). In the inset, we show a magnification of the $\alpha$ interval from 3 to 7, made in order to highlight the magnitude of the error bars. Full lines are spline interpolants. Right: The Maxwell-like construction (fluid, red crosses; solid, blue circles) allowing to determine the exact transition threshold.}
\label{fig4}
\end{figure}

Before presenting the full phase diagram we introduce an alternative method to draw the function $\Delta{\cal E}(\alpha)$ for fixed values of $R$ and $\rho$, which may also serve to check consistency with MC data. This method is fully analytic and consists in reconstructing the variational energy through the exact calculation of a sufficiently large number of $\psi$ and $\psi^2$ modes (cf. Eq.\,(\ref{a-7})). Once the latter quantities have been computed, the energy will be determined as a function of $\alpha$ through the formula (by far more compact than the sum of (\ref{eq2-8}) and (\ref{eq2-9})):
\be
{\cal E}=\frac{\hbar^2}{2mR^2}\sum_{lm}l(l+1)|c_{lm}|^2+\frac{N-1}{2}\sum_{lm}\left(2\pi\int_{-1}^1{\rm d}x\,u(x)P_l(x)\right)|d_{lm}|^2\,.
\label{eq4-3}
\ee

Let us rewrite the variational wave function (\ref{eq3-2}) as:
\be
\psi(\hat{\bf r};\alpha)=C_\alpha\sum_{k=1}^{n}\exp\left\{-D\alpha(1-\hat{\bf r}\cdot\hat{\bf R}_k)\right\}\equiv C_\alpha\sum_{k=1}^{n}h(\hat{\bf r}\cdot\hat{\bf R}_k;\alpha)\,.
\label{eq4-4}
\ee
Using the {\em Funk-Hecke formula}~\cite{Estrada},
\be
\int{\rm d}^2\Omega\,Y_l^m(\hat{\bf r})A(\hat{\bf r}\cdot\hat{\bf v})=A_lY_l^m(\hat{\bf v})\,\,\,\,\,\,{\rm with}\,\,\,\,\,\,A_l=\int{\rm d}^2\Omega\,A(\cos\theta)P_l(\cos\theta)\,,
\label{eq4-5}
\ee
which is a remarkable integral identity holding for any sufficiently regular function $A$ of $x\in[-1,1]$, we first obtain $c_{lm}$ (up to the still unknown $C_\alpha$ constant) as:
\be
c_{lm}(\alpha)=(-1)^mC_\alpha h_l(\alpha)\sum_{k=1}^{n}Y_l^{-m}(\hat{\bf R}_k)\,\,\,{\rm with}\,\,\,h_l(\alpha)=2\pi\int_{-1}^1{\rm d}x\,h(x;\alpha)P_l(x)\,.
\label{eq4-6}
\ee
As for $d_{lm}$, one observes that
\ba
\psi^2(\hat{\bf r})&=&C_\alpha^2\sum_{k,k'}\exp\left\{-D\alpha\left(2-\hat{\bf r}\cdot{\bf v}_{k,k'}\right)\right\}
\nonumber \\
&=&C_\alpha^2\sum_{k,k'}\left\{\delta_{v_{k,k'},0}e^{-2D\alpha}+(1-\delta_{v_{k,k'},0})h_{k,k'}(\hat{\bf r}\cdot\hat{\bf v}_{k,k'};\alpha)\right\}
\label{eq4-7}
\ea
with ${\bf v}_{k,k'}=\hat{\bf R}_k+\hat{\bf R}_{k'}\,,v_{k,k'}=|{\bf v}_{k,k'}|$, and $h_{k,k'}(\hat{\bf r}\cdot\hat{\bf v}_{k,k'};\alpha)=\exp\{-D\alpha\left[2-v_{k,k'}(\hat{\bf r}\cdot\hat{\bf v}_{k,k'})\right]\}$. In the double sum above, the contribution from those pairs of vertices that are diametrically opposite on the sphere has been taken into account separately. Upon inserting the above expression into the second of Eqs.\,(\ref{a-7}), we get: 
\ba
d_{lm}(\alpha)&=&C_\alpha^2\sum_{k,k'}\left\{\delta_{v_{k,k'},0}\delta_{l,0}\sqrt{4\pi}e^{-2D\alpha}+(1-\delta_{v_{k,k'},0})(-1)^mh_{k,k',l}(\alpha)Y_l^{-m}(\hat{\bf v}_{k,k'})\right\}
\label{eq4-8}
\ea
with $h_{k,k',l}(\alpha)=2\pi\int_{-1}^1{\rm d}x\,h_{k,k'}(x;\alpha)P_l(x)$. Finally, we obtain $C_\alpha$ by imposing $\psi$ normalization:
\ba
C_\alpha^{-2}&=&\sum_{k,k'}\int{\rm d}^2\Omega\left\{\delta_{v_{k,k'},0}e^{-2D\alpha}+(1-\delta_{v_{k,k'},0})h_{k,k'}(\hat{\bf r}\cdot\hat{\bf v}_{k,k'};\alpha)\right\}
\nonumber \\
&=&2\pi e^{-2D\alpha}\sum_{k,k'}\left\{\delta_{v_{k,k'},0}2+(1-\delta_{v_{k,k'},0})\frac{e^{D\alpha v_{k,k'}}-e^{-D\alpha v_{k,k'}}}{D\alpha v_{k,k'}}\right\}\,.
\label{eq4-9}
\ea
In the above equation, the integral of $h_{k,k'}$ over the full solid angle was evaluated in a coordinate system where $\hat{\bf v}_{k,k'}$ is aligned with the $z$ axis (moreover, notice that $C_0^{-2}=4\pi n^2$).

%
%
\begin{figure}
\begin{center}
\includegraphics[width=12cm]{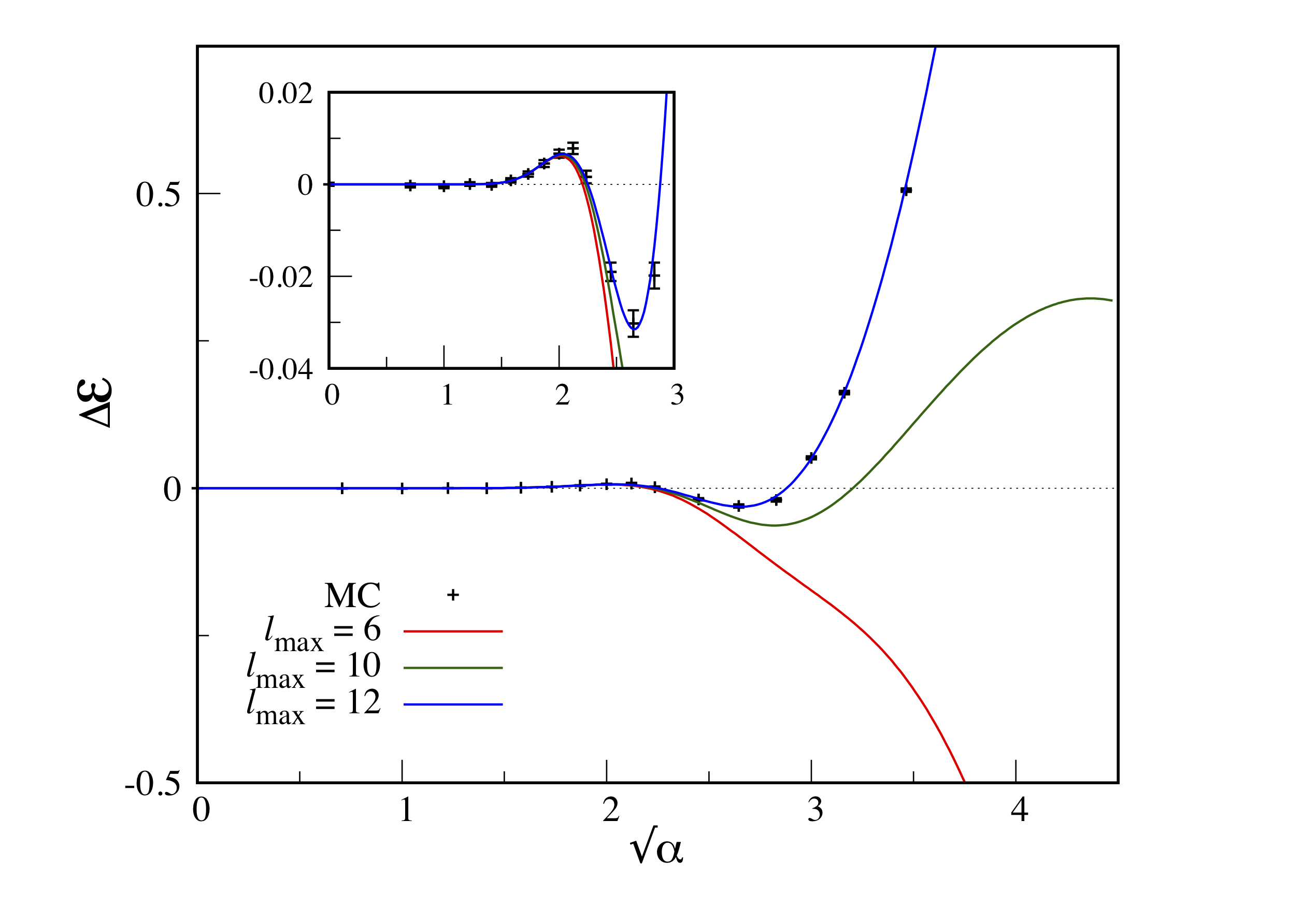}
\end{center}
\caption{PSM bosons on a sphere at $T=0$: excess energy $\Delta{\cal E}$ (units of $e_0$) of the icosahedral cluster phase, plotted as a function of $\alpha$ for $R=1.4\sigma$ and $\rho=14$ (units of $\sigma^{-2}e_0/\epsilon$). The full lines were obtained from Eq.\,(\ref{eq4-3}) by including in the two series only terms up to $l=l_{\rm max}$. The data points are results from MC integration.}
\label{fig5}
\end{figure}

Equations (\ref{eq4-6}), (\ref{eq4-7}), and (\ref{eq4-9}) allow to determine $c_{lm}$ and $d_{lm}$ exactly for all $l$ and $m$, and then the energy ${\cal E}$ from Eq.\,(\ref{eq4-3}). Apparently, this route to ${\cal E}$ has always to be preferred to MC integration. In practice, a limitation comes from the rate of convergence of the two series (\ref{a-7}), which is slower the larger $R$ and $\rho$, and this entails computing a lot of Fourier coefficients (the calculation of Legendre polynomials and spherical harmonics for large $l$ would not be a problem, as it can be carried out to any desired precision using the recurrence relations obeyed by these functions).

For the sake of clarity, let us consider the icosahedral case. For $R=1.4$ and $\rho=14$, we have computed the Fourier coefficients of $\psi$ and $\psi^2$ up to $l=16$. Giving the icosahedron an orientation such that two of its vertices lie on the $z$ axis, the only non-zero coefficients are those for $l=0,6,10,12,16$ and $m=0,\pm 5,\pm 10,\pm 15$, and are all real (a different orientation would imply different coefficients, but the weights $\sum_m|c_{lm}|^2$ and $\sum_m|d_{lm}|^2$ of each $l$ sector will be invariant~\cite{note}). By truncating {\em both} series in Eq.\,(\ref{eq4-3}) at $l_{\rm max}=6,10,12$ (which, we stress, is not equivalent to truncating the $\psi$ series (\ref{eq2-7}) at $l=l_{\rm max}$), one obtains the energy plots in Fig.\,5. We see that the MC data are already well reproduced with $l_{\rm max}=12$ (up to $\alpha\approx 20$), while a smaller $l_{\rm max}$ is insufficient to obtain good results unless $\alpha$ is low (indeed, with the exception of $l=m=0$, $|c_{lm}|$ and $|d_{lm}|$ are all increasing functions of $\alpha$).

%
%
\begin{figure}
\begin{center}
\includegraphics[width=14cm]{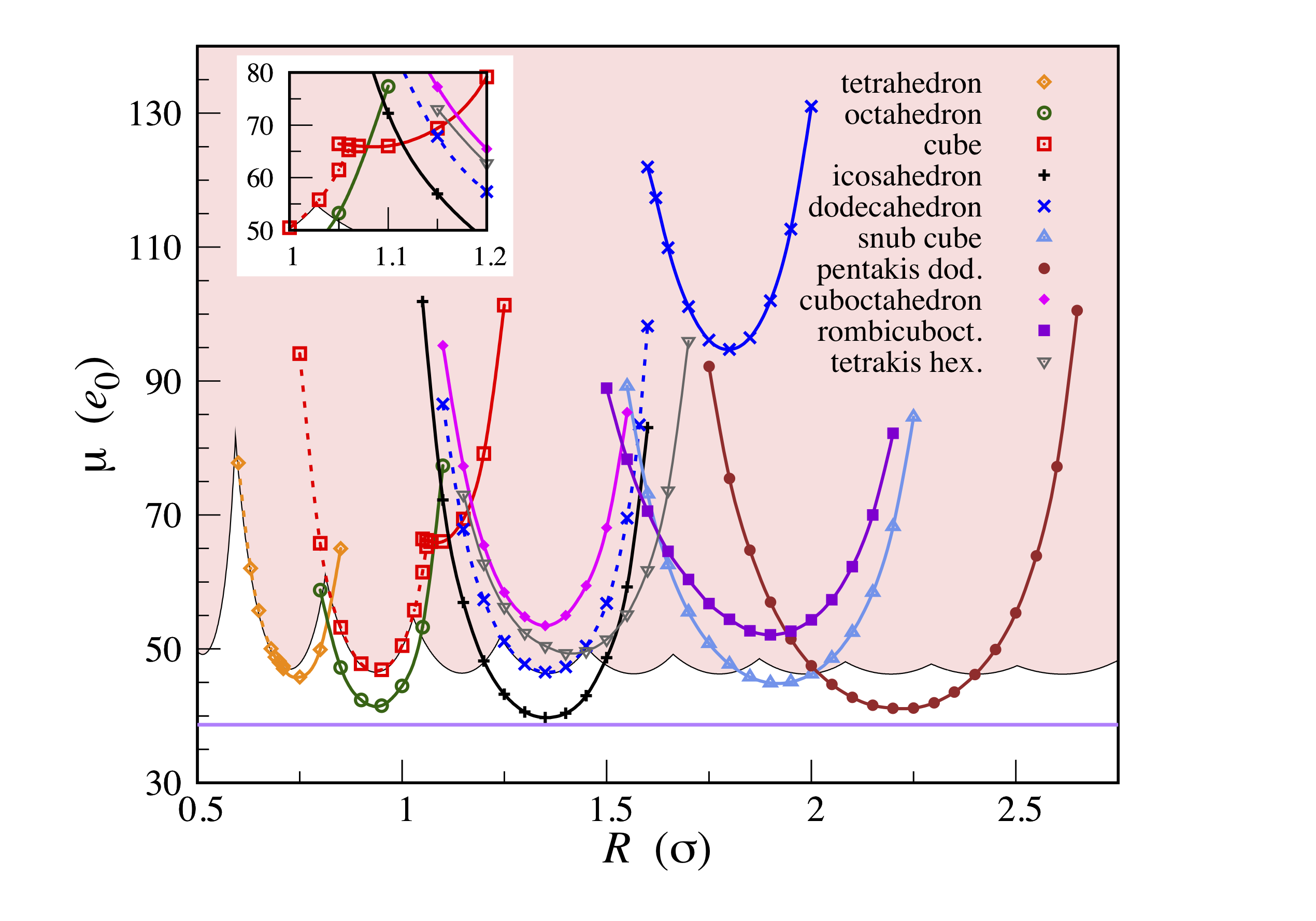}
\end{center}
\caption{PSM bosons on a sphere at $T=0$: fluid-solid transition lines according to variational theory (notice that the same wave function (\ref{eq3-2}) was also used in the two ``Catalan'' cases, despite the vertices of the latter polyhedra are of two different kinds). Continuous freezing is marked by a dashed line, whereas full lines indicate first-order freezing. The horizontal purple line marks two-dimensional freezing~\cite{Prestipino3}. The pink region is where the fluid is unstable (see text).}
\label{fig6}
\end{figure}

%
%
\begin{figure}
\begin{center}
\includegraphics[width=12cm]{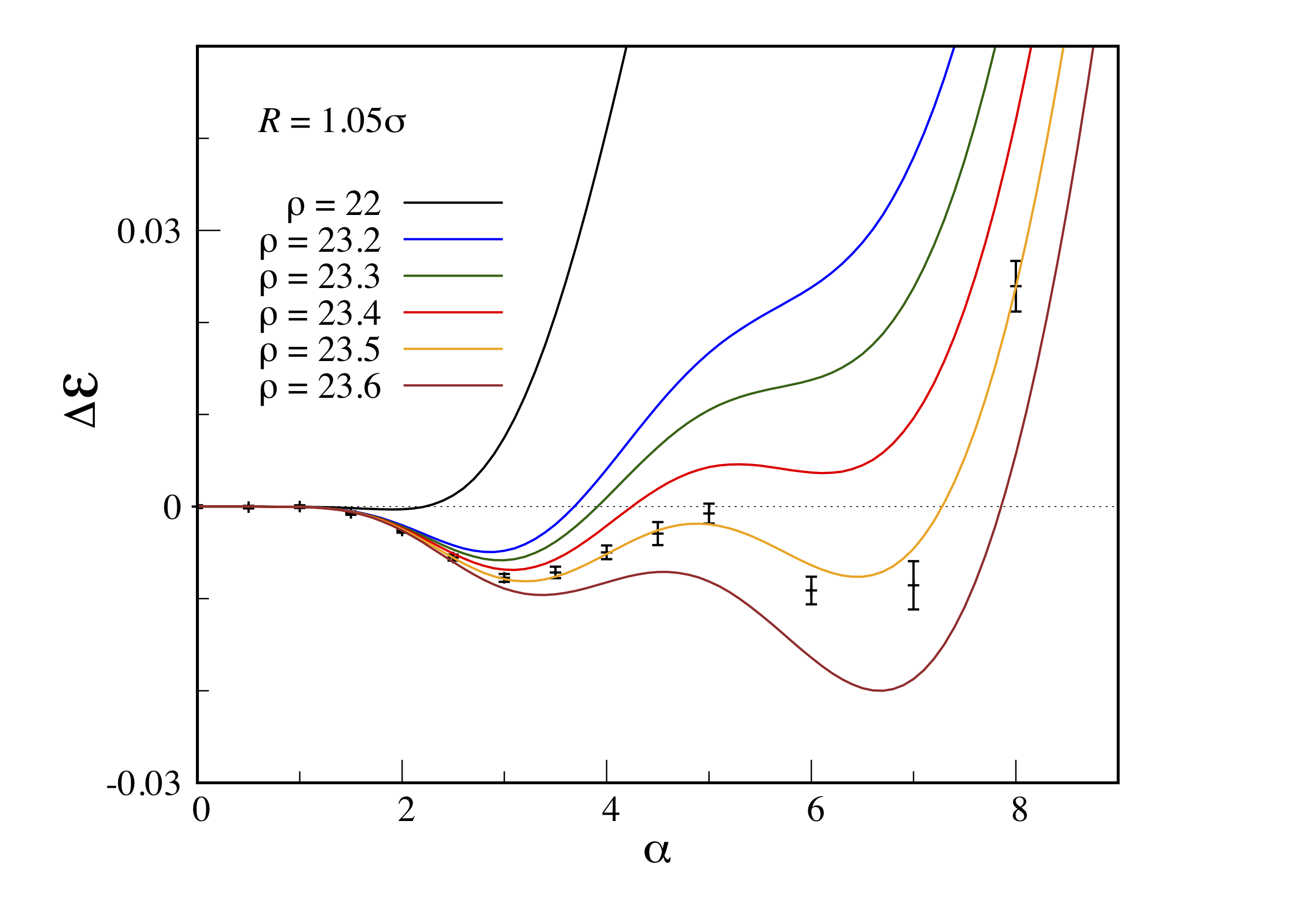}
\end{center}
\caption{PSM bosons on a sphere at $T=0$: excess energy $\Delta{\cal E}$ (units of $e_0$) of the cubic cluster phase, plotted as a function of $\alpha$ for $R=1.05\sigma$ and a number of densities in the range enclosing the two-step transition. The lines are ``exact'' results obtained from Eq.\,(\ref{eq4-3}) by including in both series all terms up to $l=16$. The data points are results from MC integration, relative to a reduced density of $\rho=23.5$ (notice the difference in energy scale between this picture and Fig.\,4).}
\label{fig7}
\end{figure}

Let us finally present our variational MF results for the ground-state diagram of spherical PSM bosons, which has been constructed by only considering the possibility of inhomogeneous phases with the symmetries of the polyhedra listed in Table I. Looking first at the fluid-solid transition lines in the $R$-$\mu$ plane (Fig.\,6), we see that each particular cluster phase can only exist in a finite range of $R$ values. In all cases, the transition occurs for a $\mu$ value larger than on the infinite plane (meaning that solid-like order is discouraged by the curvature of the sphere, as expected). We find a difference in behavior between those cases (tetrahedron, cube, and dodecahedron) where each cluster has only three other clusters around, and the other phases with a ``cluster coordination number'' $Z$ larger than 3. While for the latter phases the transition is invariably first-order, it is of mixed type for $Z=3$, i.e., continuous for small radii and first-order otherwise, with both characters coexisting in a narrow interval of radii. For a $R$ in this range, on increasing $\mu$ the fluid first freezes continuously; then, a secondary minimum develops in $\Delta{\cal E}(\alpha)$, until the system eventually undergoes a second, now discontinuous, isostructural transition (see an example in Fig.\,7 and another one in Fig.\,8). This scenario is only in part reminiscent of the behavior in flat space, where the transition is always continuous for open lattices, while being first-order for the compact ones~\cite{Prestipino3}.

%
%
\begin{figure}
\begin{center}
\includegraphics[width=12cm]{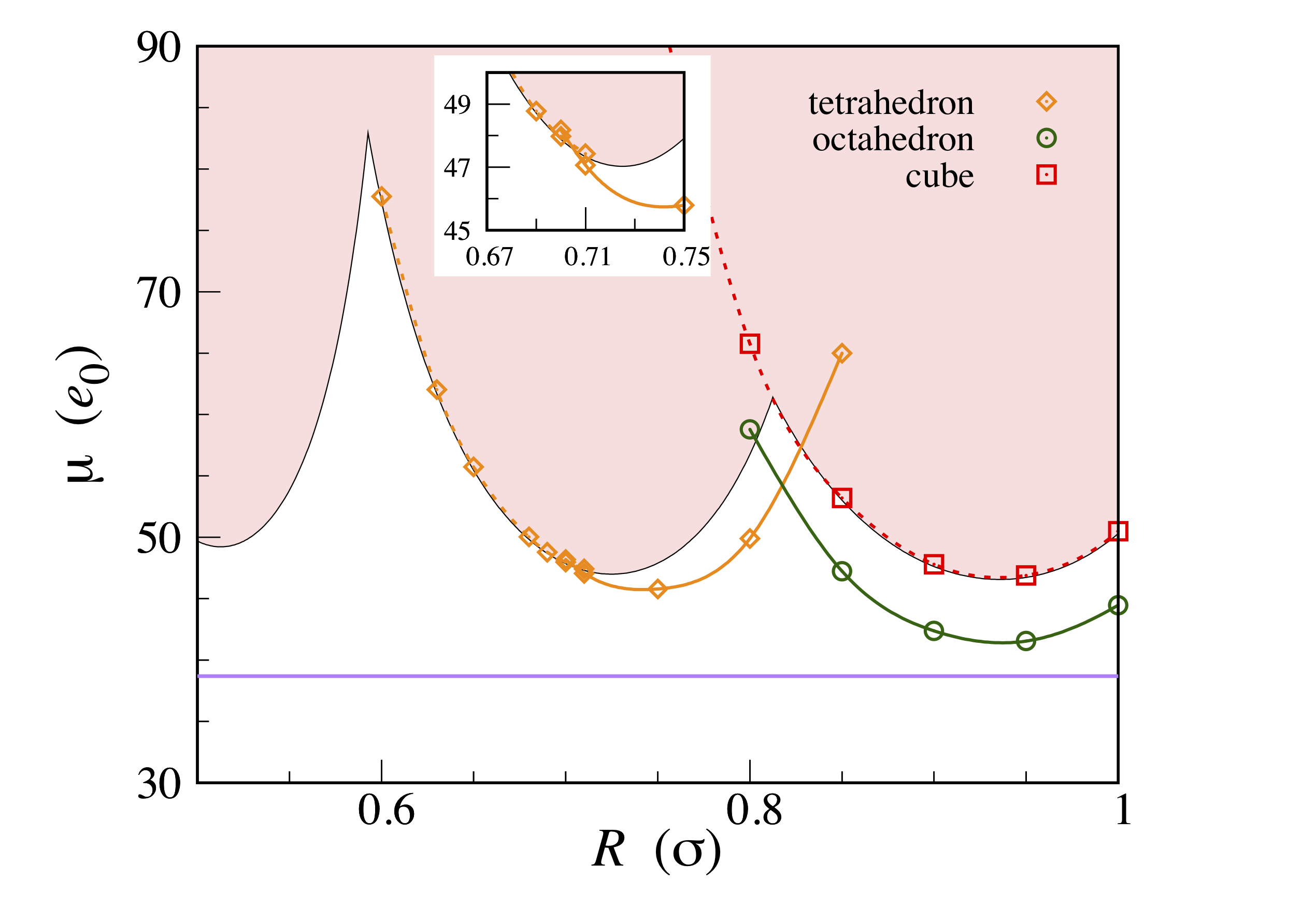}
\end{center}
\caption{PSM bosons on a sphere at $T=0$: we here highlight the region of low $R$ values, where the stable cluster phase has tetrahedral symmetry (orange). Its melting line consists of a continuous portion (dashed line) and a first-order portion (full line). There is a narrow range of radii where freezing proceeds in two steps. The emergence of tetrahedral ordering at high density is not exclusive of soft particles, since it is also found in hard particles~\cite{Prestipino9}.}
\label{fig8}
\end{figure}

%
%
\begin{figure}
\begin{center}
\includegraphics[width=14cm]{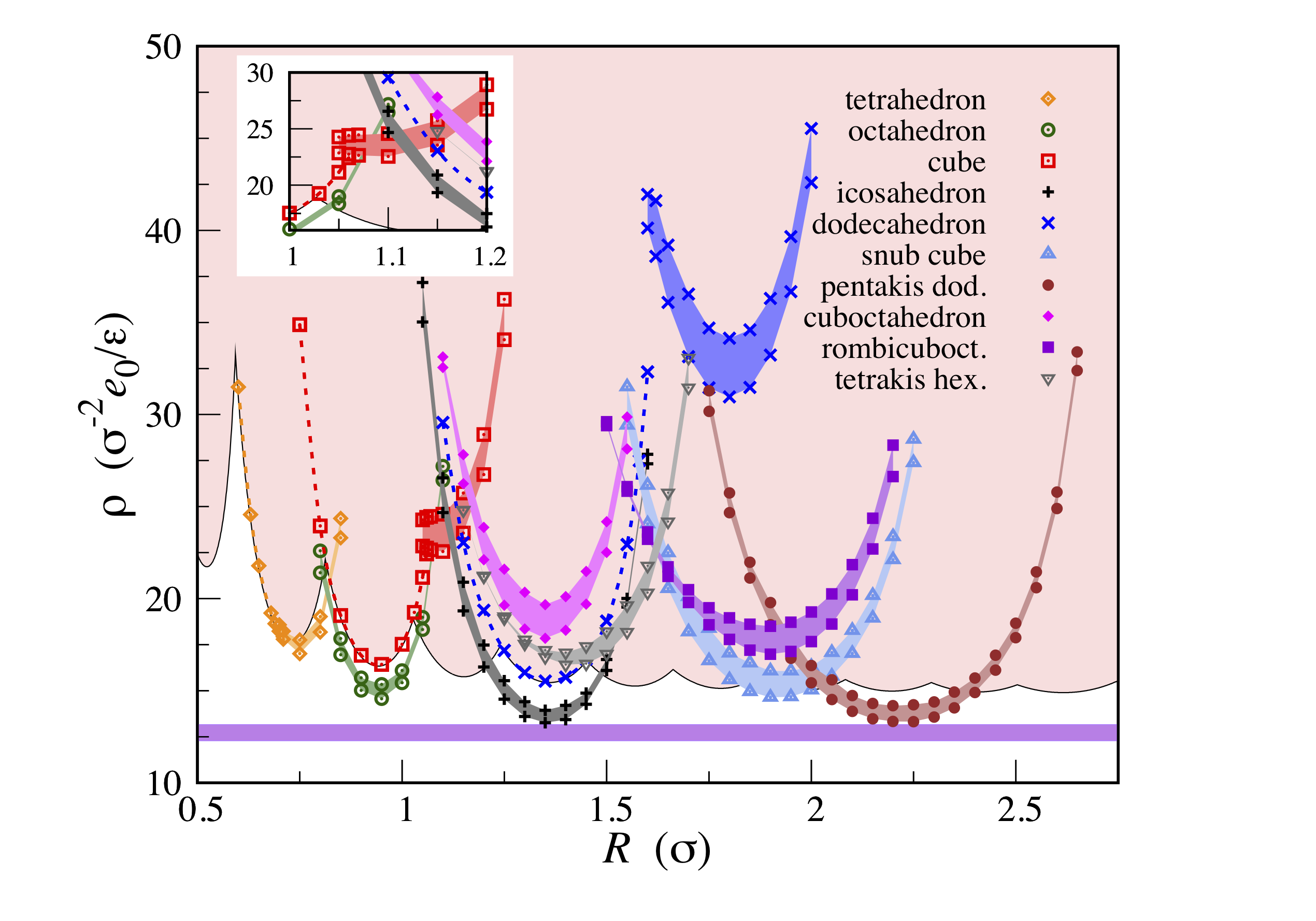}
\end{center}
\caption{PSM bosons on a sphere at $T=0$: freezing and melting lines according to variational theory. The colorful shadow regions represent fluid-solid coexistence regions, while dashed lines indicate continuous freezing. The horizontal purple stripe marks the region of coexistence between the planar fluid and the triangular cluster crystal~\cite{Prestipino3}.}
\label{fig9}
\end{figure}

%
%
\begin{figure}
\begin{center}
\includegraphics[width=8cm]{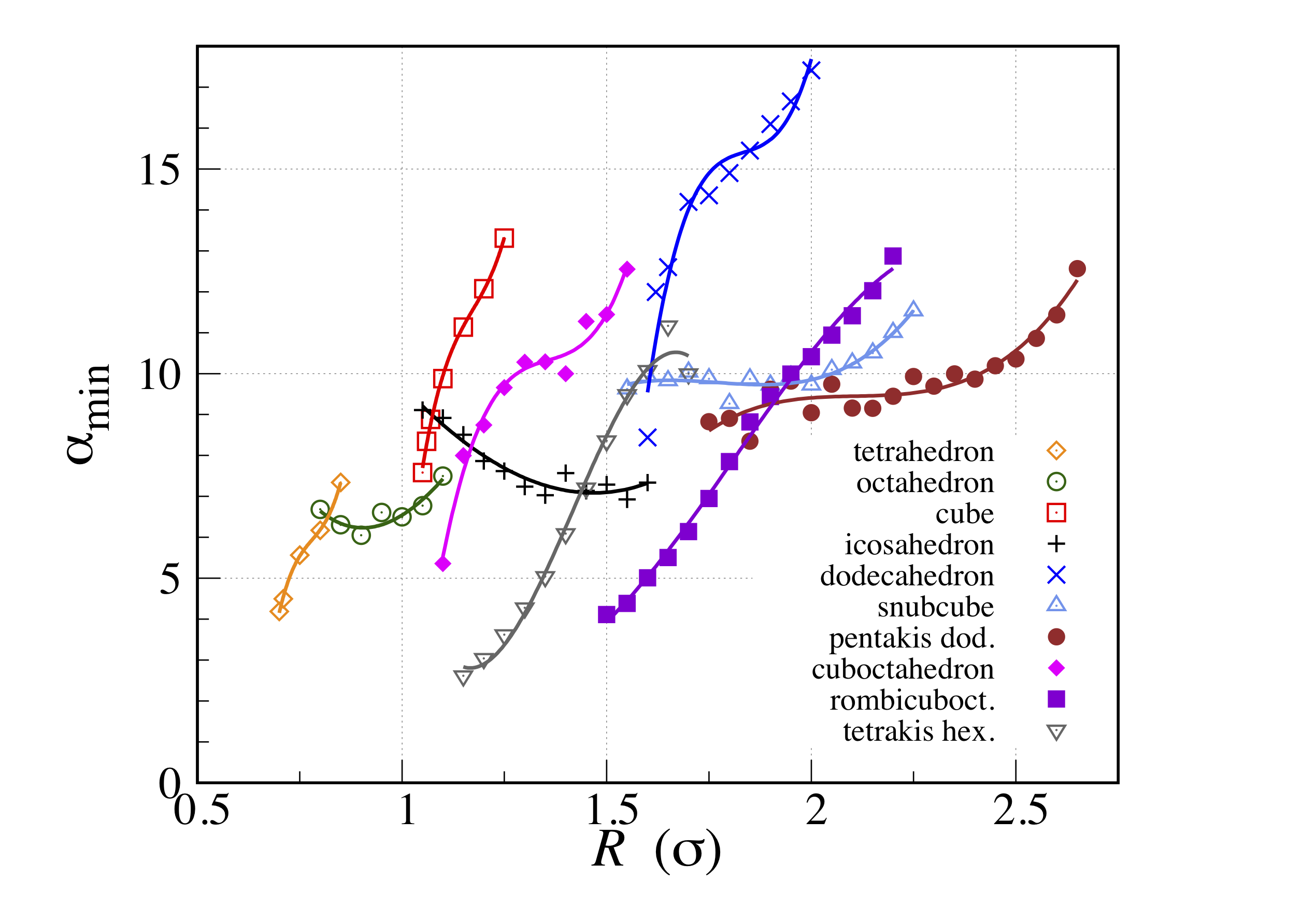}
\includegraphics[width=8cm]{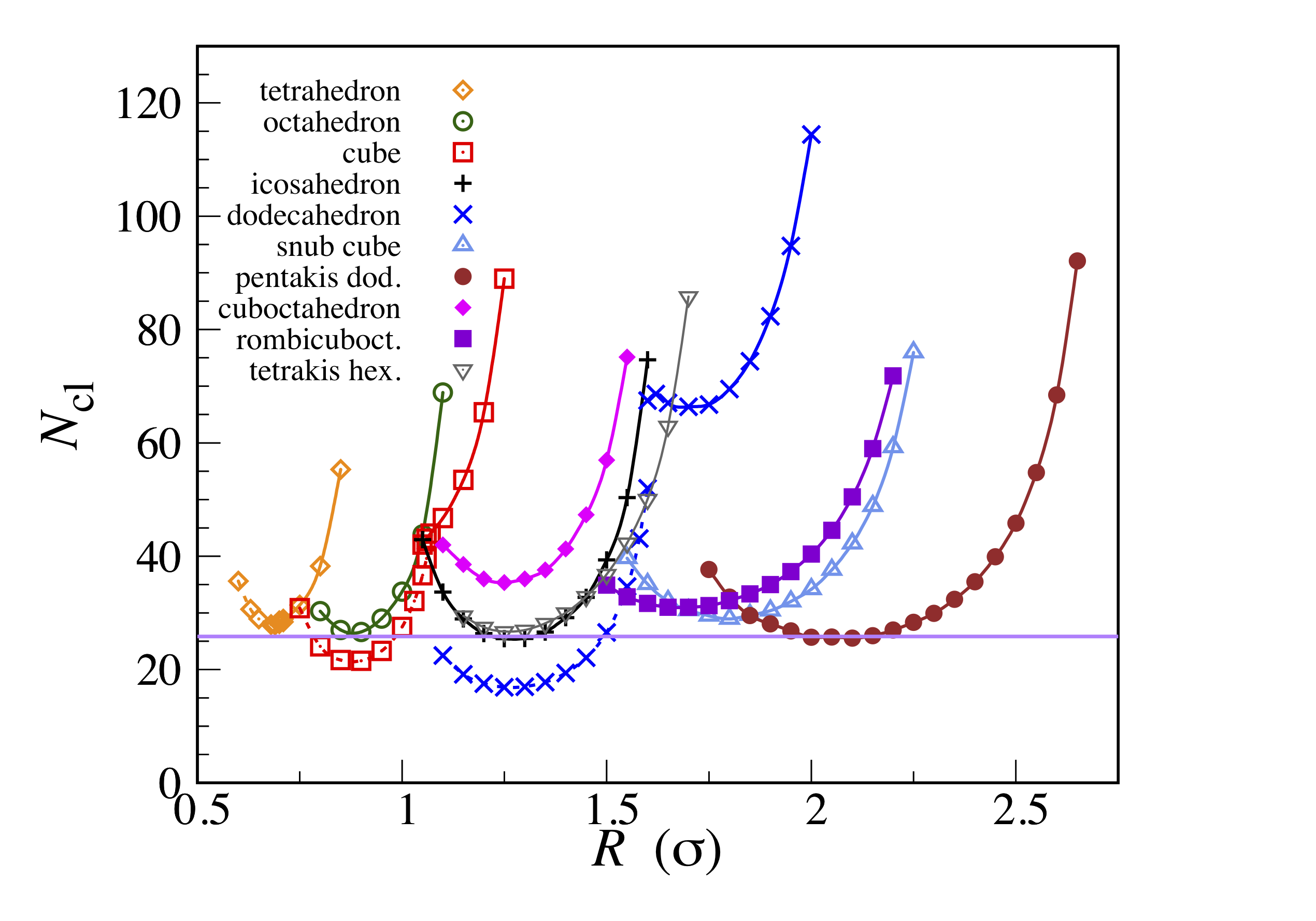}
\end{center}
\caption{PSM bosons on a sphere at $T=0$. Left: $\alpha$ value at melting for the various solid phases, plotted as a function of the radius (the full lines represent 4th-order polynomial interpolants through the data points). We stress that the imprecise determination of the location of the energy minimum only barely affects the estimate of the minimum itself. Right: average number $N_{\rm cl}$ of particles per cluster at melting. The purple straight line at $25.827$ represents the value of $N_{\rm cl}$ for the triangular cluster crystal~\cite{Prestipino3}.}
\label{fig10}
\end{figure}

In the same Fig.\,6 we have highlighted in pink the region of $R$-$\mu$ plane where the fluid is mechanically unstable. In flat space the loss of fluid stability above a certain density is heralded by the softening of roton-like excitations, which signals an instability towards the formation of a solid-like density wave. The same happens on a sphere, and we discuss at length in Appendix D how this phenomenon precisely occurs as a function of $R$. A remarkable finding is that, similarly as on a plane~\cite{Prestipino3}, continuous freezing falls exactly at the upper stability threshold of the fluid. For example, solid-like fluctuations with $l=3$ become costless right at the continuous transition to a tetrahedral phase, whose lowest non-zero modes beyond $l=0$ are indeed $l=3$ and $m=0,\pm 3$ (when one vertex of the tetrahedron lies at the north pole of the sphere). A further message from Fig.\,6 is the existence of $R$ intervals where, on increasing the density, the fluid becomes unstable {\em before} freezing. This is clearly impossible, and the reason why this occurs is that we have actually missed to identify all the relevant phases of the system --- since, probably, the underlying polyhedra have non-equivalent vertices, i.e., they have low symmetry or no regularity at all (suggestions on where to search may come from numerical studies of the Thomson problem~\cite{Wales,Neubauer}).

The freezing and melting lines of spherical PSM bosons at $T=0$ are shown in Fig.\,9. Here we can appreciate the difference in ``transition strength'' between the various phases, which not for nothing is higher for the polyhedra having large faces (i.e., for the solids whose vertices are less efficiently spread over the surface). For completeness, we report in Fig.\,10 (left panel) the position, denoted $\alpha_{\rm min}$, of the negative minimum in $\Delta{\cal E}(\alpha)$. We see that $\alpha_{\rm min}$ typically increases with $R$ and is larger the less stable the cluster phase (but there are anyway exceptions). Instead, in the right panel of Fig.\,10 we show the value at melting of the mean cluster size $N_{\rm cl}$ as a function of $R$. We see that, in the $R$ range where each cluster phase is maximally stable, $N_{\rm cl}$ lies between 20 and 30 for most of the phases, i.e., near the value ($25.83$) characteristic of the triangular cluster crystal~\cite{Prestipino3}.

%
%
\begin{figure}
\begin{center}
\includegraphics[width=12cm]{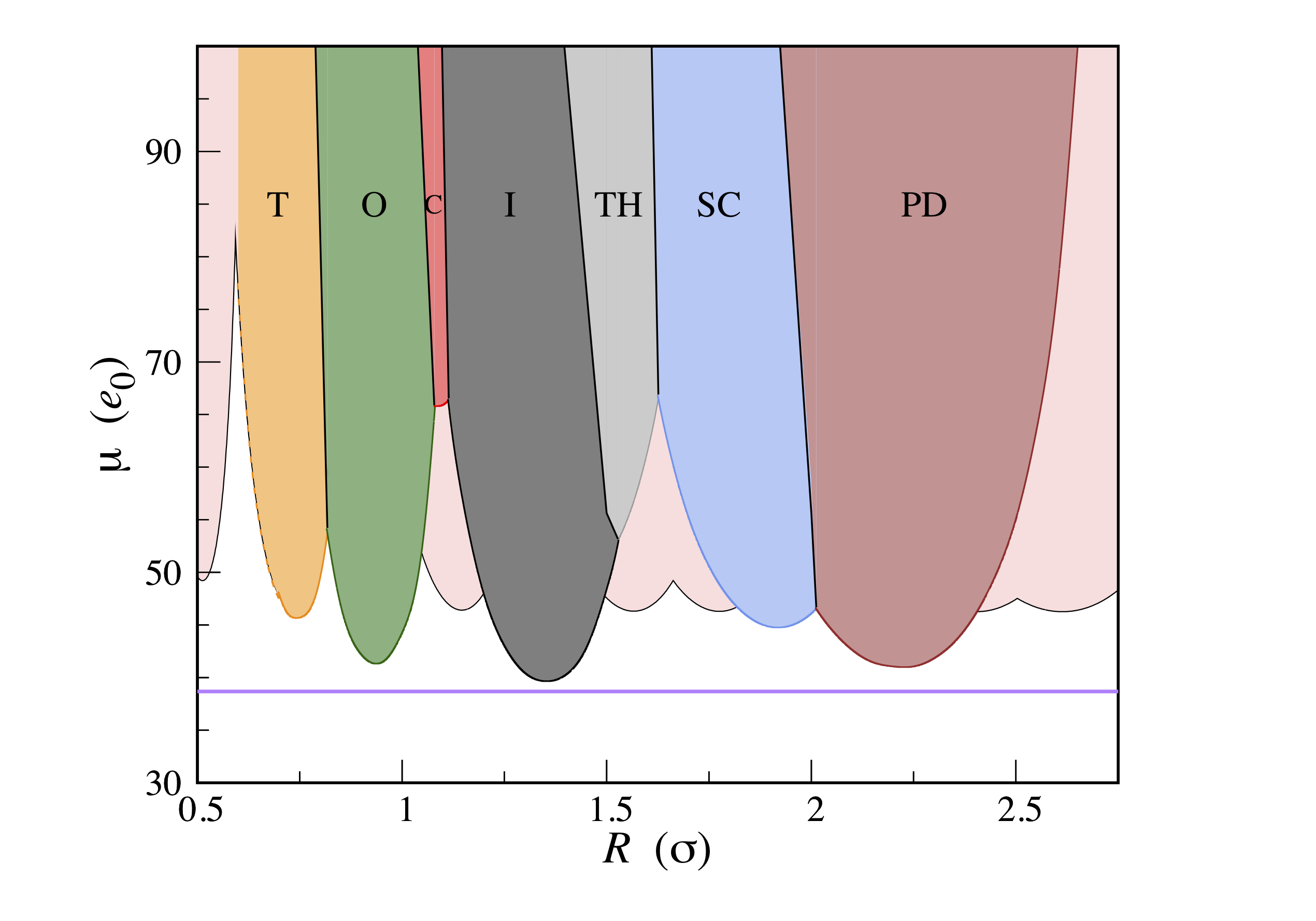}
\end{center}
\caption{PSM bosons on a sphere at $T=0$: phase diagram according to variational theory. As $R$ increases, the stable cluster phase changes accordingly, successively taking the symmetry of a tetrahedron (T), octahedron (O), cube (C), icosahedron (I), tetrakis hexahedron (TH), snub cube (SC), and pentakis dodecahedron (PD). Solid-solid lines were drawn on the basis of the few points (two or three) which we have been able to locate on iso-$R$ lines through a comparison between grand potentials. Since in the pink regions the fluid is unstable, there are ranges of $R$ where the stable high-density phase of the system is actually unknown.}
\label{fig11}
\end{figure}

Finally, the full phase diagram of the system at $T=0$ is presented in Fig.\,11. It includes as many as seven cluster phases in the $R$ interval from roughly $0.5\sigma$ to $2.5\sigma$. However, as commented before, this list of thermodynamically stable phases is far from exhaustive. Even in the quoted range of $R$, we can safely say that not all ground states of the system have been identified; on the other hand, the analysis made already allows to draw some conclusions: a) Each cluster phase spans a certain interval of $R$; if in a given range of radii there are many phases competing for stability, the winner is the one providing the most efficient occupation of the surface, or, equivalently, the highest cluster coordination number $Z$. b) As the sphere radius increases, the particles find it convenient to re-organize, adjusting the number of clusters on the surface so as to keep the distance between neighboring droplets near the magic value of $1.51\sigma$ (for the PSM); in turn, this implies an increase of $Z$ with $R$ towards the asymptotic value of 6. c) While the number of clusters is determined by $R/\sigma$, it is nevertheless nearly independent of the density (indeed, the solid-solid loci in Fig.\,11 are almost vertical).

\section{Supersolidity of the spherical cluster phases}
\setcounter{equation}{0}
\renewcommand{\thesubsection}{\arabic{subsection}}
\renewcommand{\theequation}{5.\arabic{equation}}

We conclude our analysis by showing that the cluster phases identified in the previous Section are all supersolid. In the supersolid phase of matter, still elusive in $^4$He but found in numerous lattice models~\cite{Batrouni,Wessel,Pollet} and, eventually, also observed in a quantum system with continuous ground-state degeneracy~\cite{Leonard}, the periodic density modulation typical of a solid coexists with the dissipationless flow of a superfluid.

As proposed by Leggett~\cite{Leggett2}, a supersolid can be characterized by its response to uniform axial rotations: under a slow rotation a fraction of the quantum solid may stand still, with the result that its moment of inertia is smaller than expected from classical mechanics. Leggett has called {\em superfluid fraction} of a quantum solid the quantity
\be
f_s=\frac{I_0-I}{I_0}\,,
\label{eq5-1}
\ee
where $I$ is the moment of inertia around the axis of rotation and $I_0$ its classical value. Supersolidity occurs when $f_s>0$.

When a system of $N$ particles is subject to rotation, say, around the $z$ axis, the infinitesimal change in energy due to rotation is $\omega{\rm d}L_z$, where $\omega$ is the angular velocity and $L_z={\cal O}(N)$ the $z$-component of the total angular momentum. For bosons in the product state (\ref{eq2-2}), the average $L_z$ per unit particle is given by the symmetrized expression
\be
\frac{\left\langle\Psi|L_z|\Psi\right\rangle}{N}=-\frac{i\hbar}{2}\hat{\bf z}\cdot\int{\rm d}^2\Omega\,{\bf r}\wedge\left(\psi^*\nabla\psi-\psi\nabla\psi^*\right)\,\,\,\,\,\,{\rm with}\,\,\,\,\,\,{\bf r}=R\hat{\bf r}
\label{eq5-2}
\ee
(where we are assuming that the sphere is immersed in 3D space). At $T=0$ we require that the state $\psi$ minimizes the free-energy functional ${\cal E}[\psi]-\omega\left\langle\Psi|L_z|\Psi\right\rangle/N$~\cite{Leggett3}, which for small $\omega$ is the specific energy in the absence of rotation minus $(I/N)\omega^2/2$~\cite{Aftalion}. In other words:
\be
\frac{I}{N}=-\left.\frac{\partial^2}{\partial\omega^2}{\rm min}_{\psi}\left\{{\cal E}[\psi]-\omega\frac{\left\langle\Psi|L_z|\Psi\right\rangle}{N}\right\}\right|_{\omega=0}\,.
\label{eq5-3}
\ee
When $\omega$ is non-zero the quantum state $\psi$ acquires a phase, $\Theta(\Omega)=\omega S(\Omega)+{\cal O}(\omega^2)$, whereas the square amplitude changes from $\eta_0$ (viz., the square amplitude for $\omega=0$) to $\eta=\eta_0+\omega\eta_1+{\cal O}(\omega^2)$. Inserting Eq.\,(\ref{c-1}) into (\ref{eq5-2}), we eventually obtain:
\be
\frac{\left\langle\Psi|L_z|\Psi\right\rangle}{N}=\frac{\hbar\omega}{4\pi}\int{\rm d}^2\Omega\,\eta_0\nabla S\cdot(\hat{\bf z}\wedge{\bf r})+{\cal O}(\omega^2)\,,
\label{eq5-4}
\ee
leading in turn, up to terms of order $\omega^3$, to
\be
{\cal E}[\psi]-\omega\frac{\left\langle\Psi|L_z|\Psi\right\rangle}{N}={\cal E}_0[\eta]+\frac{\hbar^2\omega^2}{8\pi m}\int{\rm d}^2\Omega\,\eta_0(\nabla S)^2-\frac{\hbar\omega^2}{4\pi}\int{\rm d}^2\Omega\,\eta_0\nabla S\cdot(\hat{\bf z}\wedge{\bf r})\,,
\label{eq5-5}
\ee
${\cal E}_0[\eta]$ being the energy functional for $\Theta={\rm const}$. Therefore:
\be
\frac{I}{N}={\rm min}_S\left\{\frac{\hbar}{2\pi}\int{\rm d}^2\Omega\,\eta_0\nabla S\cdot(\hat{\bf z}\wedge{\bf r})-\frac{\hbar^2}{4\pi m}\int{\rm d}^2\Omega\,\eta_0(\nabla S)^2\right\}\,.
\label{eq5-6}
\ee
Upon considering that
\be
\frac{I_0}{N}=\frac{1}{4\pi}\int{\rm d}^2\Omega\,\eta_0mr_\perp^2=\frac{m}{4\pi}\int{\rm d}^2\Omega\,\eta_0(\hat{\bf z}\wedge{\bf r})^2\,,
\label{eq5-7}
\ee
we finally obtain:
\be
f_s=\frac{\hbar^2}{m^2}\frac{{\rm min}_S\left\{\int{\rm d}^2\Omega\,\eta_0\left[\nabla S-(m/\hbar)\hat{\bf z}\wedge{\bf r}\right]^2\right\}}{\int{\rm d}^2\Omega\,\eta_0(\hat{\bf z}\wedge{\bf r})^2}\,.
\label{eq5-8}
\ee
While the calculation of $f_s$ is difficult, finding a lower value is easier:
\be
f_s\ge\frac{\eta_{0,{\rm min}}}{\eta_{0,{\rm max}}}\frac{\hbar^2}{m^2}\frac{{\rm min}_S\left\{\int{\rm d}^2\Omega\left[\nabla S-(m/\hbar)\hat{\bf z}\wedge{\bf r}\right]^2\right\}}{\int{\rm d}^2\Omega(\hat{\bf z}\wedge{\bf r})^2}=\frac{\eta_{0,{\rm min}}}{\eta_{0,{\rm max}}}\,,
\label{eq5-9}
\ee
where $\eta_{0,{\rm min}}$ and $\eta_{0,{\rm max}}$ are the minimum and maximum values of $\eta_0$ on the sphere. Hence, a finite density contrast $\eta_{0,{\rm max}}/\eta_{0,{\rm min}}$ is the fingerprint of supersolidity. To obtain the estimate in Eq.\,(\ref{eq5-9}), we have considered that the minimum of $\int{\rm d}^2\Omega\left[\nabla S-(m/\hbar)\hat{\bf z}\wedge{\bf r}\right]^2$ is reached for $\nabla S=0$ (indeed, the Euler-Lagrange equation for the functional in (\ref{eq5-9}) is $\nabla^2S=0$, because $\nabla\cdot(\hat{\bf z}\wedge{\bf r})=0$, and the only regular $S(\Omega)$ with $\nabla^2S=0$ is proportional to $Y_0^0$, hence it is a constant). Since $\eta_0$ is positive definite for our variational solution, we conclude that $f_s$ is strictly positive --- in other words, in our theory all cluster phases are {\em supersolid for every $\mu$}. It cannot be excluded that, for very large $\mu$, a $\psi$ function vanishing in the interstitial region between the polyhedron vertices can have a lower free energy than the Gaussian ansatz. Even in this case, supersolidity will occur at least in the vicinity of the freezing point.

\section{Conclusions}

An efficient method to study weakly-interacting bosonic particles at zero temperature is mean-field theory, which assumes a perfect condensate for the system state. As a further simplification, the single-particle wave function can be accurately modeled through some physically-motivated ansatz~\cite{Prestipino3,Prestipino4}, which is then optimized by use of the variational method. In the present study, we have employed variational mean-field theory to investigate a (finite) system of penetrable bosons confined to a spherical surface, essentially with the aim to follow the evolution of ordering tendencies with the radius $R$ in a genuinely quantum system.

The ground-state diagram of the system is very rich, featuring many distinct high-density ``phases'' as a function of $R$, all characterized by the presence of clusters of overlapping particles. The mechanism behind the onset of cluster phases on a sphere is the same as in flat space, and is purely energetic in character. We have found that many, but definitely not every, of these phases have clusters distributed at the vertices of a regular or semi-regular polyhedron inscribed in the sphere, and the stable phase at a given $R$ is the one ensuring the maximum possible (cluster) coordination number that is consistent with a distance between neighboring clusters of about $1.51\sigma$ (for PSM bosons~\cite{Prestipino3}), i.e., roughly the same as in the triangular cluster crystal. The existence of cluster phases is intimately rooted in the characteristics of the interaction between particles, and can be anticipated from the nature of the elementary excitations of the (super)fluid phase, which are roton-like for sufficiently large densities. When the roton mode eventually softens, the fluid becomes unstable towards a solid-like density modulation, and that marks the upper boundary of the homogeneous phase. In practice, unless the cluster coordination number is very small, the phase transition occurs before reaching the instability threshold, and in this case freezing is first-order.

Finally, we have given an analytic argument showing that, at least according to our variational analysis, all cluster phases are supersolid, i.e., they exhibit a reduced moment of inertia compared to its classical value. We ascribe this property to the finite strength of interparticle forces, which, by allowing particles to diffuse freely within the surface, can sustain a superfluid component in the cluster phases.

Our results can find application for the behavior of ultracold gases of bosonic atoms confined in spherically-symmetric bubble traps~\cite{Zobay,Garraway}, as will be made available in future experiments carried out in a microgravity environment~\cite{Elliott}.

\vspace{8mm}
\noindent{\bf Acknowledgements}
\vspace{5mm}

One of us (S. P.) wishes to thank A. Sergi and E. Bruno for stimulating discussions at an early stage of this research.

\appendix
\section{Gross-Pitaevskii equation on a sphere}
\renewcommand{\theequation}{A.\arabic{equation}}

In this Appendix we give a variational derivation of the GP equation, different from the one provided in \cite{Rogel-Salazar} and adapted to the sphere case.

The starting point is the MF energy functional, written for a general (i.e., normalizable but not necessarily of unit norm) single-particle wave function $\psi$:
\be
\frac{\langle\Psi|H|\Psi\rangle}{\langle\Psi|\Psi\rangle}=\langle K\rangle+\langle U\rangle
\label{a-1}
\ee
with
\be
\langle K\rangle=-N\frac{\hbar^2}{2m}\frac{\int{\rm d}^2\Omega\,\psi^*\nabla^2\psi}{\int{\rm d}^2\Omega\,\psi^*\psi}
\label{a-2}
\ee
and
\be
\langle U\rangle=\frac{N(N-1)}{2}\frac{\int{\rm d}^2\Omega\,{\rm d}^2\Omega'\,|\psi(\Omega)|^2u(\hat{\bf r}\cdot\hat{\bf r}')|\psi(\Omega')|^2}{\left(\int{\rm d}^2\Omega\,\psi^*\psi\right)^2}\,.
\label{a-3}
\ee
In Eqs.\,(\ref{a-2}) and (\ref{a-3}), the solid-angle element is ${\rm d}^2\Omega=\sin\theta\,{\rm d}\theta{\rm d}\phi$. According to the variational principle, the ``best'' approximate ground state is such that the average energy $\langle H\rangle$ in (\ref{a-1}) is minimum. For the latter to occur, a necessary condition is $\delta\langle H\rangle=0$. Upon observing that
\ba
\frac{\delta\langle K\rangle}{\delta\psi^*(\Omega)}&=&-N\frac{\hbar^2}{2m}\frac{\nabla^2\psi\int{\rm d}^2\Omega'\,\psi^*\psi-\psi\int{\rm d}^2\Omega'\,\psi^*\nabla^2\psi}{\left(\int{\rm d}^2\Omega'\,\psi^*\psi\right)^2}\,;
\nonumber \\
\frac{\delta\langle U\rangle}{\delta\psi^*(\Omega)}&=&N(N-1)\frac{\psi\left(\int{\rm d}^2\Omega'\,\psi^*\psi\right)^2\int{\rm d}^2\Omega'\,\psi^*u\psi-\psi\int{\rm d}^2\Omega'\,\psi^*\psi\int{\rm d}^2\Omega'\,{\rm d}^2\Omega''\,|\psi|^2u|\psi|^2}{\left(\int{\rm d}^2\Omega'\,\psi^*\psi\right)^4}\,,
\nonumber \\
\label{a-4}
\ea
we readily arrive at:
\be
-\frac{\hbar^2}{2m}\nabla^2\psi+(N-1)\psi(\Omega)\frac{\int{\rm d}^2\Omega'\,|\psi(\Omega')|^2u(\hat{\bf r}\cdot\hat{\bf r}')}{\int{\rm d}^2\Omega'\,\psi^*\psi}=\lambda\psi(\Omega)
\label{a-5}
\ee
with
\be
\lambda=-\frac{\hbar^2}{2m}\frac{\int{\rm d}^2\Omega\,\psi^*\nabla^2\psi}{\int{\rm d}^2\Omega\,\psi^*\psi}+(N-1)\frac{\int{\rm d}^2\Omega\,{\rm d}^2\Omega'\,|\psi(\Omega)|^2u(\hat{\bf r}\cdot\hat{\bf r}')|\psi(\Omega')|^2}{\left(\int{\rm d}^2\Omega\,\psi^*\psi\right)^2}\,.
\label{a-6}
\ee
It is hardly necessary to observe that, when the radius $R$ is of order $\sigma$, $N-1$ may not be quite the same as $N$. Now observe that, if $\psi$ obeys (\ref{a-5}) also $c\psi$ is a solution, for all $c\ne 0$. In particular, we can always choose $c$ such that $\psi$ is normalized to 1. In this case, Eq.\,(\ref{a-5}) becomes the GP equation in its standard form (see Eqs.\,(\ref{eq2-3}) and (\ref{eq2-5})), with $\lambda=(\langle K\rangle+2\langle U\rangle)/N$.  

The most natural way to solve the GP equation is to expand the solution and its square modulus in a series of spherical harmonics,
\be
\psi(\Omega)=\sum_{lm}c_{lm}Y_l^m(\Omega)\,\,\,\,\,\,{\rm and}\,\,\,\,\,\,|\psi(\Omega)|^2=\sum_{lm}d_{lm}Y_l^m(\Omega)\,,
\label{a-7}
\ee
where the $d_{lm}$ coefficients can clearly be expressed in terms of the $c_{lm}$ themselves (see Eq.\,(\ref{a-9}) below). Observe that the $d_{lm}$ do not fulfil any particular sum rule (because, at variance with $\psi$, the function $|\psi|^2$ is not subject to any specific normalization). Moreover, since $|\psi(\Omega)|^2$ is real, it is generally $d_{l,-m}=(-1)^md_{lm}^*$. The rest of the derivation follows the one provided in Appendix B for the MF energy functional, and we finally arrive at:
\ba
&&\frac{\hbar^2}{2mR^2}l(l+1)c_{lm}+(N-1)\sum_{l'm'}c_{l'm'}\sum_{l_3m_3}\left(2\pi\int_{-1}^1{\rm d}x\,u(x)P_{l_3}(x)\right)d_{l_3m_3}
\nonumber \\
&\times&(-1)^m\left(\begin{array}{ccc}l & l' & l_3 \\0 & 0 & 0\end{array}\right)\left(\begin{array}{ccc}l & l' & l_3 \\-m & m' & m_3\end{array}\right)\sqrt{\frac{(2l+1)(2l'+1)(2l_3+1)}{4\pi}}=\lambda c_{lm}
\label{a-8}
\ea
with
\ba
d_{l_3m_3}&=&\sum_{l_1,m_1,l_2,m_2}(-1)^{m_2+m_3}c_{l_1m_1}c_{l_2m_2}^*
\nonumber \\
&\times&\left(\begin{array}{ccc}l_1 & l_2 & l_3 \\0 & 0 & 0\end{array}\right)\left(\begin{array}{ccc}l_1 & l_2 & l_3 \\m_1 & -m_2 & -m_3\end{array}\right)\sqrt{\frac{(2l_1+1)(2l_2+1)(2l_3+1)}{4\pi}}\,.
\label{a-9}
\ea
Equation (\ref{a-8}) is a non-linear set of equations, similar to that reported in Eq.\,(2.6) of Ref.\,\cite{Prestipino3}. Therefore, it is tempting to solve it self-consistently, assuming arbitrary initial values for the $c_{lm}$ and filtering out the minimum energy eigenvector of unit norm at each iteration step. However, while this method works well in the planar case it dramatically fails to converge in the present case (unless the density is very small), and the problem is not alleviated by the use of a mixing scheme. To guarantee that iteration of (\ref{a-8}) becomes a contractive fixed point iteration, we may think to replace spherical harmonics with a different basis of functions, possibly a different one for any specific radius and polyhedral symmetry. However, besides the difficulty of devising specific basis functions for each case, the price to pay is losing the good properties of spherical harmonics that allow simplifying the final form of the GP equation. In view of this, a more viable procedure is to resort to a variational ansatz for $\psi$, which also offers the advantage of better elucidating the physics behind the minimum-energy state.

\section{Proof of Eq.\,(\ref{eq2-9})}
\renewcommand{\theequation}{B.\arabic{equation}}

Here we obtain an expression for the potential-energy term in Eq.\,(\ref{eq2-6}). We first observe that any bounded $u$ with finite support can be written as a Fourier integral:
\be
u(x)=\int_{-\infty}^{+\infty}\frac{{\rm d}k}{2\pi}\,\widetilde{u}(k)e^{ikx}\,\,\,\,\,\,{\rm with}\,\,\,\,\,\,\widetilde{u}^*(k)=\widetilde{u}(-k)
\label{b-1}
\ee
(notice that $k$ is a dimensionless variable). Upon considering that
\be
Y_l^{m*}(\hat{\bf r})=(-1)^mY_l^{-m}(\hat{\bf r})\,\,\,\,\,\,{\rm and}\,\,\,\,\,\,Y_l^m(-\hat{\bf r})=(-1)^lY_l^m(\hat{\bf r})\,,
\label{b-2}
\ee
and using the expansion of a plane wave in spherical harmonics, we get:
\ba
u(\hat{\bf r}\cdot\hat{\bf r}')&=&\int_{-\infty}^{+\infty}\frac{{\rm d}k}{2\pi}\,\widetilde{u}(k)e^{ik\hat{\bf r}\cdot\hat{\bf r}'}=\int_0^{+\infty}\frac{{\rm d}k}{2\pi}\left[\widetilde{u}(-k)e^{-ik\hat{\bf r}\cdot\hat{\bf r}'}+\widetilde{u}(k)e^{ik\hat{\bf r}\cdot\hat{\bf r}'}\right]
\nonumber \\
&=&2\sum_{lm}i^l(-1)^mY_l^m(\hat{\bf r})Y_l^{-m}(\hat{\bf r}')\int_0^{+\infty}{\rm d}k\left[(-1)^l\widetilde{u}(-k)+\widetilde{u}(k)\right]j_l(k)\,,
\label{b-3}
\ea
$j_l(k)$ being a spherical Bessel function. Alternatively, and also more conveniently, $u(x)$ can be written as a series of Legendre polynomials:
\be
u(x)=\sum_{l=0}^\infty\left(\frac{2l+1}{2}\int_{-1}^1{\rm d}t\,u(t)P_l(t)\right)P_l(x)\,.
\label{b-4}
\ee
By noting that
\be
P_l(\hat{\bf r}\cdot\hat{\bf r}')=\frac{4\pi}{2l+1}\sum_{m=-l}^lY_l^m(\hat{\bf r})Y_l^{m*}(\hat{\bf r}')\,,
\label{b-5}
\ee
we promptly obtain:
\be
u(\hat{\bf r}\cdot\hat{\bf r}')=\sum_{lm}(-1)^m\left(2\pi\int_{-1}^1{\rm d}x\,u(x)P_l(x)\right)Y_l^m(\hat{\bf r})Y_l^{-m}(\hat{\bf r}')\,.
\label{b-6}
\ee
In the following we prefer using (\ref{b-6}) in Eq.\,(\ref{eq2-6}), rather than (\ref{b-3}), while deferring a direct proof of
\be
2i^l\int_0^{+\infty}{\rm d}k\left[(-1)^l\widetilde{u}(-k)+\widetilde{u}(k)\right]j_l(k)=2\pi\int_{-1}^1{\rm d}x\,u(x)P_l(x)
\label{b-7}
\ee
to below in this Appendix.

Next, we note that
\be
|\psi(\Omega)|^2=\sum_{lm,l'm'}c_{l'm'}^*c_{lm}Y_{l'}^{m'*}(\Omega)Y_l^m(\Omega)=\sum_{lm,l'm'}(-1)^{m'}c_{l'm'}^*c_{lm}Y_{l'}^{-m'}(\Omega)Y_l^m(\Omega)\,.
\label{b-8}
\ee
Putting all things together:
\ba
{\cal E}_{\rm pot}&=&\frac{N-1}{2}\int{\rm d}^2\Omega\,{\rm d}^2\Omega'\sum_{l_1m_1,l_2m_2}(-1)^{m_2}c_{l_1m_1}c_{l_2m_2}^*Y_{l_1}^{m_1}(\hat{\bf r})Y_{l_2}^{-m_2}(\hat{\bf r})
\nonumber \\
&\times&\sum_{lm}(-1)^m\left(2\pi\int_{-1}^1{\rm d}x\,u(x)P_l(x)\right)Y_l^m(\hat{\bf r})Y_l^{-m}(\hat{\bf r}')
\nonumber \\
&\times&\sum_{l_3m_3,l_4m_4}(-1)^{m_4}c_{l_3m_3}c_{l_4m_4}^*Y_{l_3}^{m_3}(\hat{\bf r}')Y_{l_4}^{-m_4}(\hat{\bf r}')\,.
\label{b-9}
\ea
The integral of a product of three spherical harmonics (also called {\em Gaunt coefficient}) has a known value:
{\small
\be
\int{\rm d}^2\Omega\,Y_{l_1}^{m_1}(\Omega)Y_{l_2}^{m_2}(\Omega)Y_{l_3}^{m_3}(\Omega)=\sqrt{\frac{(2l_1+1)(2l_2+1)(2l_3+1)}{4\pi}}\left(\begin{array}{ccc}l_1 & l_2 & l_3 \\0 & 0 & 0\end{array}\right)\left(\begin{array}{ccc}l_1 & l_2 & l_3 \\m_1 & m_2 & m_3\end{array}\right)\,,
\label{b-10}
\ee
}
where the $2\times 3$ tables are Wigner 3-j symbols~\cite{Stone}. As a result, Eq.\,(\ref{b-9}) gets simplified, eventually transforming into Eq.\,(\ref{eq2-9}).

Finally, we provide a proof of Eq.\,(\ref{b-7}), which establishes the equivalence between (\ref{b-3}) and (\ref{b-6}). We seek for a different expression of
\be
\int_0^{+\infty}{\rm d}k\left[(-1)^l\widetilde{u}(-k)+\widetilde{u}(k)\right]j_{l}(k)\,.
\label{b-11}
\ee
Taking advantage of the formula\,\cite{Ludu}
\be
j_l(k)=\frac{1}{2}(-i)^l\int_{-1}^1{\rm d}t\,e^{ikt}P_l(t)\,,
\label{b-12}
\ee
which is valid for all $l$ and $k$, we first obtain (for any $x$ satisfying $-1<x<1$):
\be
\int_{-\infty}^{+\infty}{\rm d}k\,e^{ikx}j_l(k)=\frac{1}{2}(-i)^l\int_{-1}^1{\rm d}t\,P_l(t)\underbrace{\int_{-\infty}^{+\infty}{\rm d}k\,e^{i(x+t)k}}_{2\pi\delta(x+t)}=\pi(-i)^lP_l(-x)=\pi i^lP_l(x)\,,
\label{b-13}
\ee
where in the last step we used the property $P_l(-x)=(-1)^lP_l(x)$. Similarly,
\be
\int_{-\infty}^{+\infty}{\rm d}k\,e^{-ikx}j_l(k)=\pi(-i)^lP_l(x)\,.
\label{b-14}
\ee
The case of $x=\pm 1$ needs a different treatment, since the delta argument vanishes at one of the extrema of the $t$ integration interval. Indeed, one can prove that
\be
\int_{-\infty}^{+\infty}{\rm d}k\,e^{\pm ik}j_l(k)=\frac{\pi}{2}(\pm i)^l
\label{b-15}
\ee
(rather than $\pi(\pm i)^l$, as it would follow from Eq.\,(\ref{b-13}) and (\ref{b-14})).

Coming to the calculation of (\ref{b-11}), we have:
\ba
\int_0^{+\infty}{\rm d}k\left[(-1)^l\widetilde{u}(-k)+\widetilde{u}(k)\right]j_{l}(k)&=&\int_{-1}^1{\rm d}x\,u(x)\int_0^{+\infty}{\rm d}k\,j_l(k)\left[(-1)^le^{ikx}+e^{-ikx}\right]
\nonumber \\
&=&\frac{1}{2}\int_{-1}^1{\rm d}x\,u(x)\int_{-\infty}^{+\infty}{\rm d}k\,j_l(k)\left[(-1)^le^{ikx}+e^{-ikx}\right]\,,
\nonumber \\
\label{b-16}
\ea
where, in consideration of $j_l(-k)=(-1)^lj_l(k)$, the last step holds for both even and odd $l$. The inner integral in (\ref{b-16}) can be evaluated for any $-1<x<1$ using Eqs.\,(\ref{b-13}) and (\ref{b-14}):
\ba
\frac{1}{2}\int_{-\infty}^{+\infty}{\rm d}k\,j_l(k)\left[(-1)^le^{ikx}+e^{-ikx}\right]=(-1)^l\int_{-\infty}^{+\infty}{\rm d}k\,j_l(k)e^{ikx}=(-1)^l\pi i^lP_l(x)\,.
\label{b-17}
\ea
Despite the case $x=\pm 1$ would need a separate treatment, the value of the integrand at the $x$ boundary in (\ref{b-16}) is irrelevant for the value of the same integral. By plugging Eq.\,(\ref{b-17}) in (\ref{b-16}), we readily arrive at the desired Eq.\,(\ref{b-7}).

\section{The condensate of minimum energy is real}
\renewcommand{\theequation}{C.\arabic{equation}}

In this Appendix, we show that the true ground state of a system of spherical bosons with specific energy given as in Eq.\,(\ref{eq2-6}) is necessarily represented by a real wave function.

Using the Madelung representation,
\be
\psi=\frac{1}{\sqrt{4\pi}}\sqrt{\eta(\Omega)}e^{i\Theta(\Omega)}
\label{c-1}
\ee
with $\eta\ge 0$ and a real $\Theta$ defined up to an arbitrary additive constant, the potential-energy term in Eq.\,(\ref{eq2-6}) immediately reads (pulling out the homogeneous-system energy):
{\small
\ba
{\cal E}_{\rm pot}&=&\frac{N-1}{2}\int{\rm d}^2\Omega\,{\rm d}^2\Omega'\,\frac{\eta(\Omega)}{4\pi}u(\hat{\bf r}\cdot\hat{\bf r}')\frac{\eta(\Omega')}{4\pi}
\nonumber \\
&=&\frac{N-1}{4}\int_{-1}^1{\rm d}x\,u(x)+\frac{N-1}{32\pi^2}\int{\rm d}^2\Omega\,{\rm d}^2\Omega'\left(\eta(\Omega)-1\right)u(\hat{\bf r}\cdot\hat{\bf r}')\left(\eta(\Omega')-1\right)\,.
\label{c-2}
\ea
}
Aside from a constant factor, the kinetic-energy term is given by
\ba
\int{\rm d}^2\Omega\,\psi^*\nabla^2\psi&=&\frac{1}{R^2}\int_0^{2\pi}{\rm d}\phi\int_0^{\pi}{\rm d}\theta\left[\psi^*\frac{\partial}{\partial\theta}\left(\sin\theta\frac{\partial\psi}{\partial\theta}\right)+\frac{\psi^*}{\sin\theta}\frac{\partial^2\psi}{\partial\phi^2}\right]
\nonumber \\
&=&-\frac{1}{R^2}\int{\rm d}^2\Omega\left(\frac{\partial\psi^*}{\partial\theta}\frac{\partial\psi}{\partial\theta}+\frac{1}{\sin^2\theta}\frac{\partial\psi^*}{\partial\phi}\frac{\partial\psi}{\partial\phi}\right)\,,
\label{c-3}
\ea
where the last step follows after partial integration. Upon inserting (\ref{c-1}) into (\ref{c-3}), and taking account of the expression
\be
\nabla f=\frac{1}{R}\frac{\partial f}{\partial\theta}\hat{\boldsymbol\theta}+\frac{1}{R\sin\theta}\frac{\partial f}{\partial\phi}\hat{\boldsymbol\phi}
\label{c-4}
\ee
of the gradient of a scalar function $f(\theta,\phi)$, we obtain:
{\footnotesize
\ba
\int{\rm d}^2\Omega\,\psi^*\nabla^2\psi&=&-\frac{1}{4\pi R^2}\int{\rm d}^2\Omega\left\{\left(\frac{\partial\sqrt{\eta}}{\partial\theta}\right)^2+\frac{1}{\sin^2\theta}\left(\frac{\partial\sqrt{\eta}}{\partial\phi}\right)^2+\eta\left[\left(\frac{\partial\Theta}{\partial\theta}\right)^2+\frac{1}{\sin^2\theta}\left(\frac{\partial\Theta}{\partial\phi}\right)^2\right]\right\}
\nonumber \\
&=&-\frac{1}{4\pi}\int{\rm d}^2\Omega\left[\left(\nabla\sqrt{\eta}\right)^2+\eta\left(\nabla\Theta\right)^2\right]\,,
\label{c-5}
\ea
}
whence finally:
\be
{\cal E}_{\rm kin}=-\frac{\hbar^2}{2m}\int{\rm d}^2\Omega\,\psi^*\nabla^2\psi=\frac{\hbar^2}{32\pi m}\int{\rm d}^2\Omega\left[\frac{(\nabla\eta)^2}{\eta}+4\eta(\nabla\Theta)^2\right]\,.
\label{c-6}
\ee
Equation (\ref{c-6}) represents, {\em mutatis mutandis} (i.e., with $4\pi$ in place of the volume), the same result holding in flat space. Looking at Eqs.\,(\ref{c-2}) and (\ref{c-6}), it is clear that the minimum (kinetic) energy is attained for a constant $\Theta$. As a global phase in the wave function cannot affect the results, we can always choose $\Theta=0$. This implies that the MF ground state is real and non-negative.

\section{Elementary excitations of the fluid}
\renewcommand{\theequation}{D.\arabic{equation}}

We hereafter investigate the collective excitations of the system in the fluid phase, using a method similar to that employed in Ref.\,\cite{Prestipino3}. The starting point is the time-dependent GP equation,
\be
i\hbar\frac{\partial\psi}{\partial t}(\Omega,t)=-\frac{\hbar^2}{2m}\nabla^2\psi(\Omega,t)+(N-1)\int{\rm d}^2\Omega'\,|\psi(\Omega',t)|^2u(\hat{\bf r}\cdot\hat{\bf r}')\psi(\Omega,t)\,,
\label{d-1}
\ee
describing the MF quantum dynamics of identical bosons at $T=0$. Multiplying Eq.\,(\ref{d-1}) by $\psi^*$ and subtracting the complex conjugate of the resulting equation, we arrive at the following continuity equation:
\be
\frac{\partial}{\partial t}(\psi^*\psi)+\frac{i\hbar}{2m}\nabla\cdot(\psi\nabla\psi^*-\psi^*\nabla\psi)=0\,,
\label{d-2}
\ee
where we have considered that the divergence of a vector field ${\bf A}=A_\theta\hat{\boldsymbol\theta}+A_\phi\hat{\boldsymbol\phi}$ is
\be
\nabla\cdot{\bf A}=\frac{1}{R\sin\theta}\left[\frac{\partial}{\partial\theta}\left(A_\theta\sin\theta\right)+\frac{\partial A_\phi}{\partial\phi}\right]\,.
\label{d-3}
\ee
In (\ref{d-2}), the velocity field is clearly
\be
{\bf v}=\frac{i\hbar}{2m}\frac{\psi\nabla\psi^*-\psi^*\nabla\psi}{|\psi|^2}=\frac{\hbar}{m}\nabla\Theta\,,
\label{d-4}
\ee
where the last step follows after inserting the Madelung form (\ref{c-1}).

Another equation connecting $\eta$ and $\Theta$ is obtained by plugging Eq.\,(\ref{c-1}) in the time-dependent GP equation, with the result that:
\ba
&&i\hbar\frac{1}{2\sqrt{\eta}}\frac{\partial\eta}{\partial t}-\hbar\sqrt{\eta}\frac{\partial\Theta}{\partial t}=-\frac{\hbar^2}{2m}\frac{i}{\sqrt{\eta}}\nabla\eta\cdot\nabla\Theta-\frac{\hbar^2}{2m}i\sqrt{\eta}\nabla^2\Theta
\nonumber \\
&+&\frac{\hbar^2}{8m}\frac{(\nabla\eta)^2}{\eta^{3/2}}-\frac{\hbar^2}{4m}\frac{\nabla^2\eta}{\sqrt{\eta}}+\frac{\hbar^2}{2m}\sqrt{\eta}(\nabla\Theta)^2+\frac{N-1}{4\pi}\int{\rm d}^2\Omega'\,\eta(\Omega',t)u(\hat{\bf r}\cdot\hat{\bf r}')\sqrt{\eta}\,.
\label{d-5}
\ea
While the imaginary part of (\ref{d-5}) gives back the continuity equation, the real part reads:
\be
-\hbar\frac{\partial\Theta}{\partial t}=-\frac{\hbar^2}{2m}\frac{\nabla^2\sqrt{\eta}}{\sqrt{\eta}}+\frac{\hbar^2}{2m}(\nabla\Theta)^2+\frac{N-1}{4\pi}\int{\rm d}^2\Omega'\,\eta(\Omega',t)u(\hat{\bf r}\cdot\hat{\bf r}')\,.
\label{d-6}
\ee
Taking the gradient of (\ref{d-6}), we arrive at a Navier-Stokes-like equation without viscosity term:
\be
m\frac{\partial{\bf v}}{\partial t}+m({\bf v}\cdot\nabla){\bf v}=\frac{\hbar^2}{2m}\nabla\left(\frac{\nabla^2\sqrt{\eta}}{\sqrt{\eta}}\right)-\frac{N-1}{4\pi}\nabla\int{\rm d}^2\Omega'\,\eta(\Omega',t)u(\hat{\bf r}\cdot\hat{\bf r}')\,,
\label{d-7}
\ee
where it is intended that
\be
({\bf A}\cdot\nabla){\bf A}=A_\theta\nabla A_\theta+A_\phi\nabla A_\phi\,.
\label{d-8}
\ee

We now derive an approximate equation valid for a $\psi$ function departing only slightly from the homogeneous-system solution, $\eta=1$ and $\nabla\Theta=0$. Such perturbed solutions are the elementary excitations of the fluid phase. Inserting $\eta=1+\delta\eta$ and $\nabla\Theta=\delta{\bf u}$ into the continuity equation (\ref{d-2}), and simply ignoring every term that is not linear in $\delta\eta$ or $\delta{\bf u}$, we first get:
\be
\frac{\partial\delta\eta}{\partial t}+\frac{\hbar}{m}\nabla\cdot\delta{\bf u}=0\,\,\,\,\,\,\Longrightarrow\,\,\,\,\,\,\frac{\partial^2\delta\eta}{\partial t^2}=-\frac{\hbar}{m}\nabla\cdot\left(\frac{\partial\delta{\bf u}}{\partial t}\right)\,.
\label{d-9}
\ee
Moreover, we have:
\ba
m\frac{\partial{\bf v}}{\partial t}&=&\hbar\frac{\partial\delta{\bf u}}{\partial t}\,;
\nonumber \\
m({\bf v}\cdot\nabla){\bf v}&=&\frac{m}{2}\nabla(v^2)=\frac{\hbar^2}{2m}\nabla(\delta u^2)={\cal O}(\delta u^2)\,;
\nonumber \\
\frac{\hbar^2}{2m}\nabla\left(\frac{\nabla^2\sqrt{\eta}}{\sqrt{\eta}}\right)&=&\frac{\hbar^2}{4m}\nabla(\nabla^2\delta\eta)\,;
\nonumber \\
-\frac{N-1}{4\pi}\nabla\int{\rm d}^2\Omega'\,\eta(\Omega',t)u(\hat{\bf r}\cdot\hat{\bf r}')&=&-\frac{N-1}{4\pi}\nabla\int{\rm d}^2\Omega'\,\delta\eta(\Omega',t)u(\hat{\bf r}\cdot\hat{\bf r}')\,,
\label{d-10}
\ea
which eventually allow us to simplify Eq.\,(\ref{d-7}) as:
\be
\hbar\frac{\partial\delta{\bf u}}{\partial t}=\frac{\hbar^2}{4m}\nabla(\nabla^2\delta\eta)-\frac{N-1}{4\pi}\nabla\int{\rm d}^2\Omega'\,\delta\eta(\Omega',t)u(\hat{\bf r}\cdot\hat{\bf r}')\,.
\label{d-11}
\ee
Inserting Eq.\,(\ref{d-11}) into the second of Eqs.\,(\ref{d-9}), we finally arrive at:
\be
\frac{\partial^2\delta\eta}{\partial t^2}=-\frac{\hbar^2}{4m^2}\nabla^2(\nabla^2\delta\eta)+\frac{N-1}{4\pi m}\nabla^2\int{\rm d}^2\Omega'\,\delta\eta(\Omega',t)u(\hat{\bf r}\cdot\hat{\bf r}')\,.
\label{d-12}
\ee
This equation admits solutions in the form $\delta\eta=\varepsilon{\rm Re}\left\{Y_l^m(\Omega)e^{i\omega_l t}\right\}$, where $\varepsilon$ is a small dimensionless amplitude. The dispersion relation of these waves can be obtained by observing that:
{\small
\ba
&&\frac{\partial^2\delta\eta}{\partial t^2}=-\omega_l^2\delta\eta\,,\,\,\,\nabla^2\delta\eta=-\frac{l(l+1)}{R^2}\delta\eta\,,\,\,\,\nabla^2(\nabla^2\delta\eta)=\left(\frac{l(l+1)}{R^2}\right)^2\delta\eta\,,\,\,\,{\rm and}
\nonumber \\
&&\nabla^2\int{\rm d}^2\Omega'\,\delta\eta(\Omega',t)u(\hat{\bf r}\cdot\hat{\bf r}')=-\frac{l(l+1)}{R^2}u_l\delta\eta(\Omega,t)\,\,\,\,\,\,\left({\rm with}\,\,\,u_l=2\pi\int_{-1}^1{\rm d}x\,u(x)P_l(x)\right)\,.
\nonumber \\
\label{d-13}
\ea
}
In particular, we derived the last equation above from the Funk-Hecke formula (\ref{eq4-5}) for $A=u$. Substituting Eqs.\,(\ref{d-13}) into (\ref{d-12}), we finally obtain:
\be
\hbar^2\omega_l^2=\frac{\hbar^2}{2m}\,\frac{l(l+1)}{R^2}\left(\frac{\hbar^2}{2m}\,\frac{l(l+1)}{R^2}+\frac{N-1}{2\pi}u_l\right)\,,
\label{d-14}
\ee
which can be viewed as the spherical version of the Bogoliubov spectrum.

%
%
\begin{figure}
\begin{center}
\includegraphics[width=12cm]{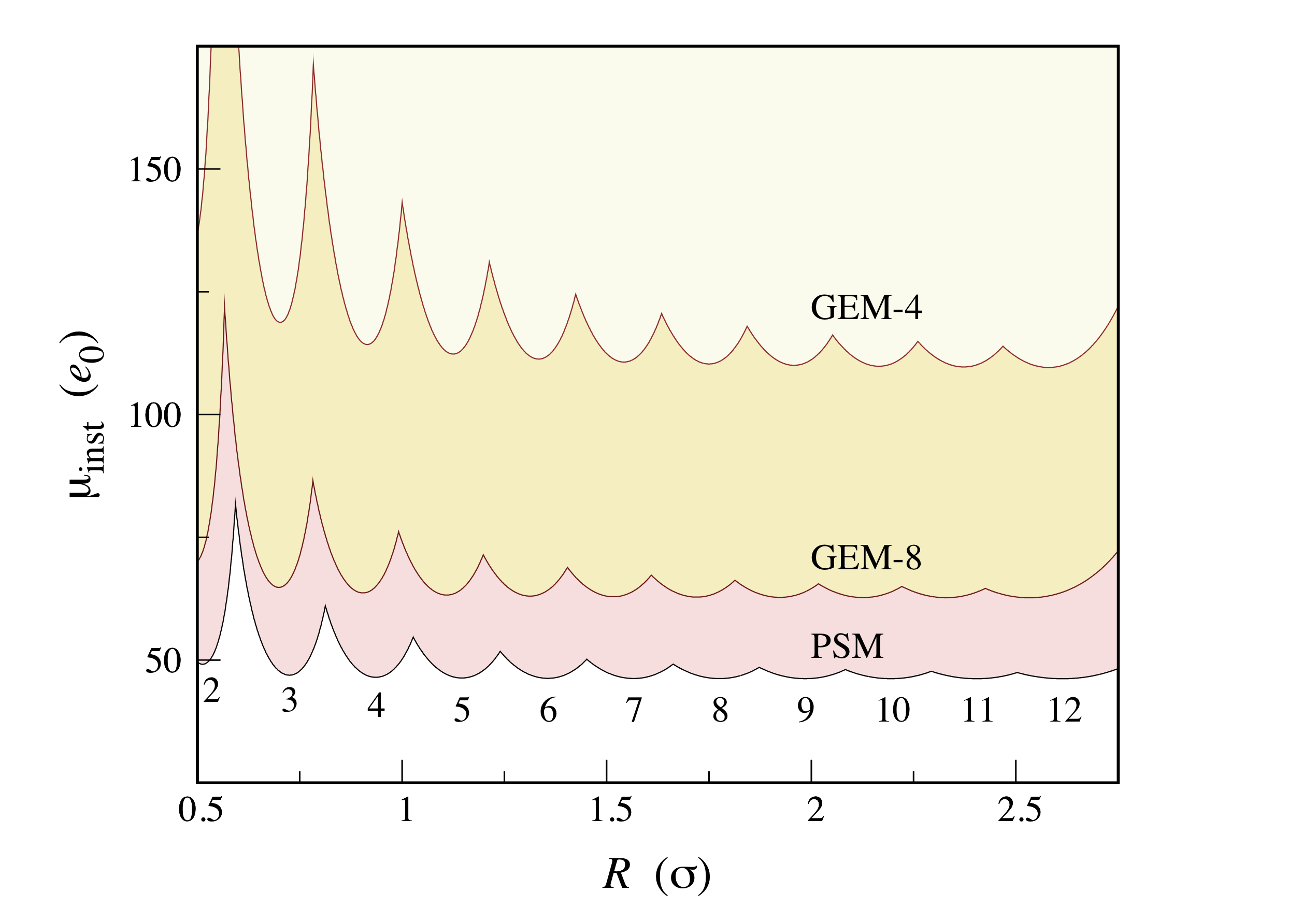}
\end{center}
\caption{Profile of $\mu_{\rm inst}(R)$, representing the upper stability threshold of the fluid phase, for three distinct interactions (PSM, GEM-8, and GEM-4). In preparing this figure, only $l$ values from 2 to 12 were considered; for each $l$, a specific minimum arises in $\mu_{\rm inst}(R)$, as marked below the graph for the PSM case.}
\label{fig12}
\end{figure}

As long as the r.h.s. of (\ref{d-14}) is positive, the fluid phase is stable and (by the same argument exposed in Sec.\,V) {\em superfluid}; conversely, if $\omega_l$ is purely imaginary, quantum dynamics will drive the system arbitrarily far from $\eta=1$. It turns out that, for each fixed $l$, $\omega_l^2$ turns from positive to negative at a density of
\be
\rho_l(R)=\frac{1}{4\pi R^2}\left[1-\frac{\pi\hbar^2}{mu_l}\,\frac{l(l+1)}{R^2}\right]\,.
\label{d-15}
\ee
Hence, when the density exceeds a certain value the fluid becomes destabilized. The upper threshold for fluid stability in terms of chemical potential is
\be
\mu_{\rm inst}(R)=\min_l\left\{\frac{4\pi R^2\rho_l-1/2}{4\pi}u_0\right\}\,.
\label{d-16}
\ee
It turns out that $\mu_{\rm inst}(R)$ shows cusps where the $l$ value providing the minimum in (\ref{d-16}) jumps by one (see Fig.\,12). Quite remarkably, the oscillatory behavior of $\mu_{\rm inst}(R)$ is similar to that computed within density-functional theory for the $\lambda$-line of a classical fluid of spherical soft-core particles~\cite{Franzini}. In Fig.\,12 the $\mu_{\rm inst}(R)$ locus is reported for three distinct models of interaction, namely the PSM, GEM-8, and GEM-4 potentials. We see that the region of fluid stability extends more and more the smoother is the generalized-exponential interaction (same as found in flat space~\cite{Prestipino3}). As $R$ grows to infinity, the instability line flattens out until it finally equates the planar threshold (e.g., $46.2979\ldots$ for the PSM potential~\cite{Prestipino3}).

\end{document}